\documentclass[aps, pra, reprint, superscriptaddress, showpacs]{revtex4-1}

\usepackage{gensymb}
\usepackage{graphicx}
\usepackage{color}
\usepackage{booktabs}
\usepackage{bm}
\usepackage{amsmath}
\usepackage{dcolumn}
\usepackage{placeins}
\usepackage{expdlist}
\usepackage{physics}
\usepackage{siunitx}
\usepackage{caption}
\usepackage{subcaption}
\usepackage{appendix}

\begin{document}

\newcommand{\avg}[1]{\langle #1 \rangle}
\newcommand{\up}{\uparrow}
\newcommand{\down}{\downarrow}
\newcommand{\eff}[1]{#1_{\rm eff}}
\renewcommand{\Re}{\mbox{Re}}
\renewcommand{\Im}{\mbox{Im}}


\title{Polarisation of radio-frequency magnetic fields in magnetic induction measurements with an atomic magnetometer}

\author{L. M. Rushton}
\author{L. M. Ellis}
\author{J. D. Zipfel}
\author{P. Bevington}
\email{patrick.bevington@npl.co.uk}
\author{W. Chalupczak}
\affiliation{National Physical Laboratory, Hampton Road, Teddington, TW11 0LW, United Kingdom}

\begin{abstract}
We explore properties of the radio-frequency atomic magnetometer, specifically its sensitivity to the polarisation of an oscillating magnetic field. This aspect can be particularly relevant to configurations where the sensor monitors fields created by more than one source. The discussion, illustrated by theoretical and experimental studies, is done in the context of the signals produced by  electrically conductive and magnetically permeable plates in magnetic induction tomography measurements. We show that different components of the secondary magnetic fields create the object response depending on the properties of the material, with the polarisation of the rf field varying across the object's surface. We argue that the ability of the sensor to simultaneously detect different field components enables the optimisation of measurement strategies for different object compositions.
\end{abstract}

\maketitle


\section{Introduction}

Since their first realization, tunable radio-frequency (rf) atomic magnetometers attracted significant attention due to their wide range of potential applications \cite{Savukov2005, Ledbetter2007, Keder2014, zigdon2010, ingleby2018, dhombridge2022, rushton2023}. A list of proof-of-principle demonstrations include nuclear quadrupole resonance \cite{Lee2006}, nuclear magnetic resonance \cite{{Savukov2007}}, magnetic induction tomography (MIT) \cite{Deans2016, Bevington2018, Bevington2019, rushton_2022, wickenbrock_2016} and, more recently, low rf communications \cite{Bevington2020d, Ingleby2020}. Most of these are active measurements, where the initial excitation - referred to as the ``primary rf field'' in the context of MIT - produces an object response, called the ``secondary rf field'', that is measured by the rf sensor.
Historically, pick-up coils have been the preferred sensor for rf field detection. One of the advantages of coil-based systems is their simplicity, however, their sensitivity is proportional to operating frequency and area (Faraday's law), which limits the performance at low frequencies. The advent of chip-scale atomic devices paved the way for atomic magnetometers to become an instrument for commercial applications \cite{Kitching2018}. 

\begin{figure*}[!htb]
\begin{center}
  \begin{tabular}{c c c} 
         \includegraphics[height=0.45\columnwidth]{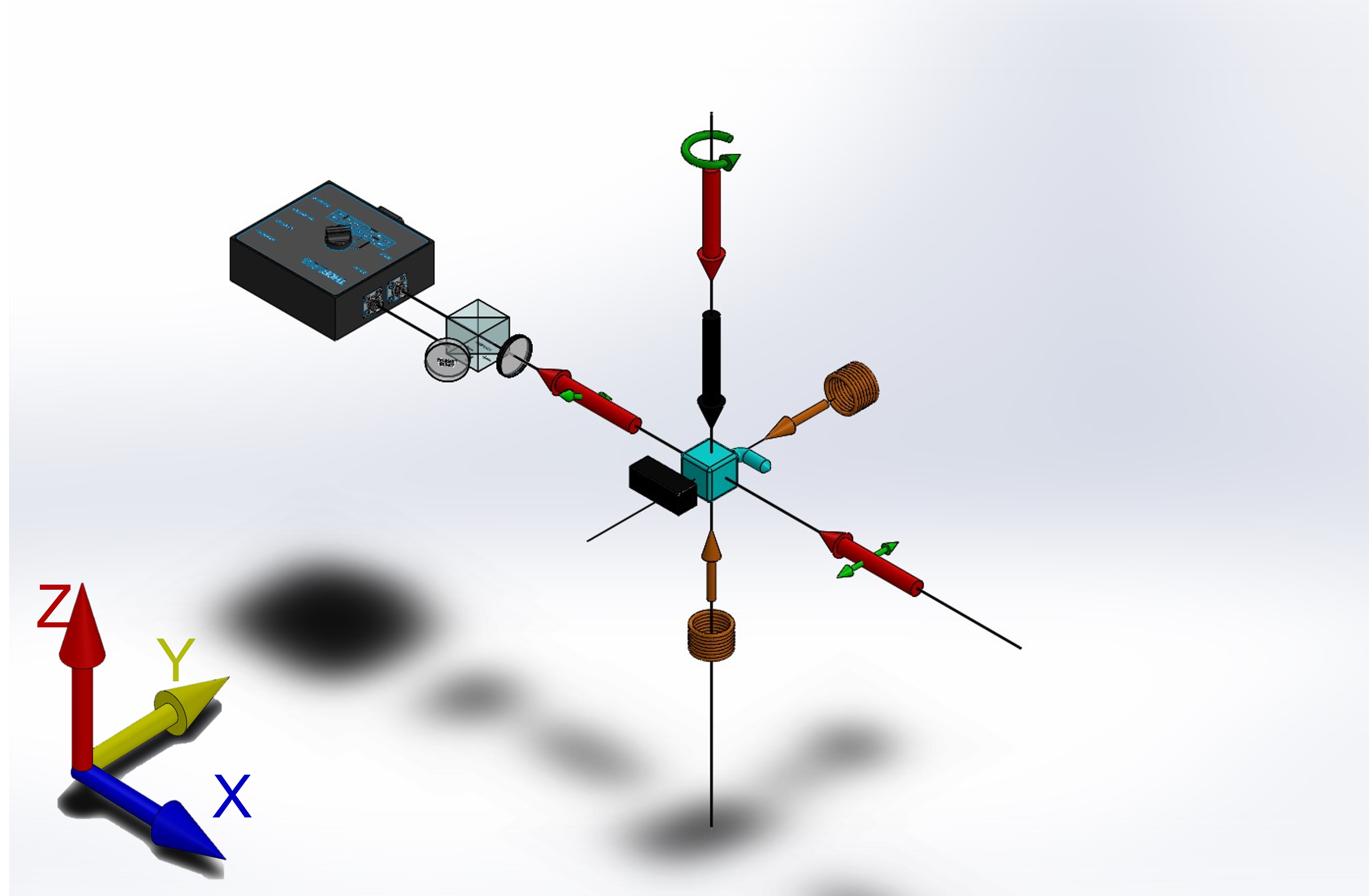} &    
         \includegraphics[height=0.45\columnwidth]{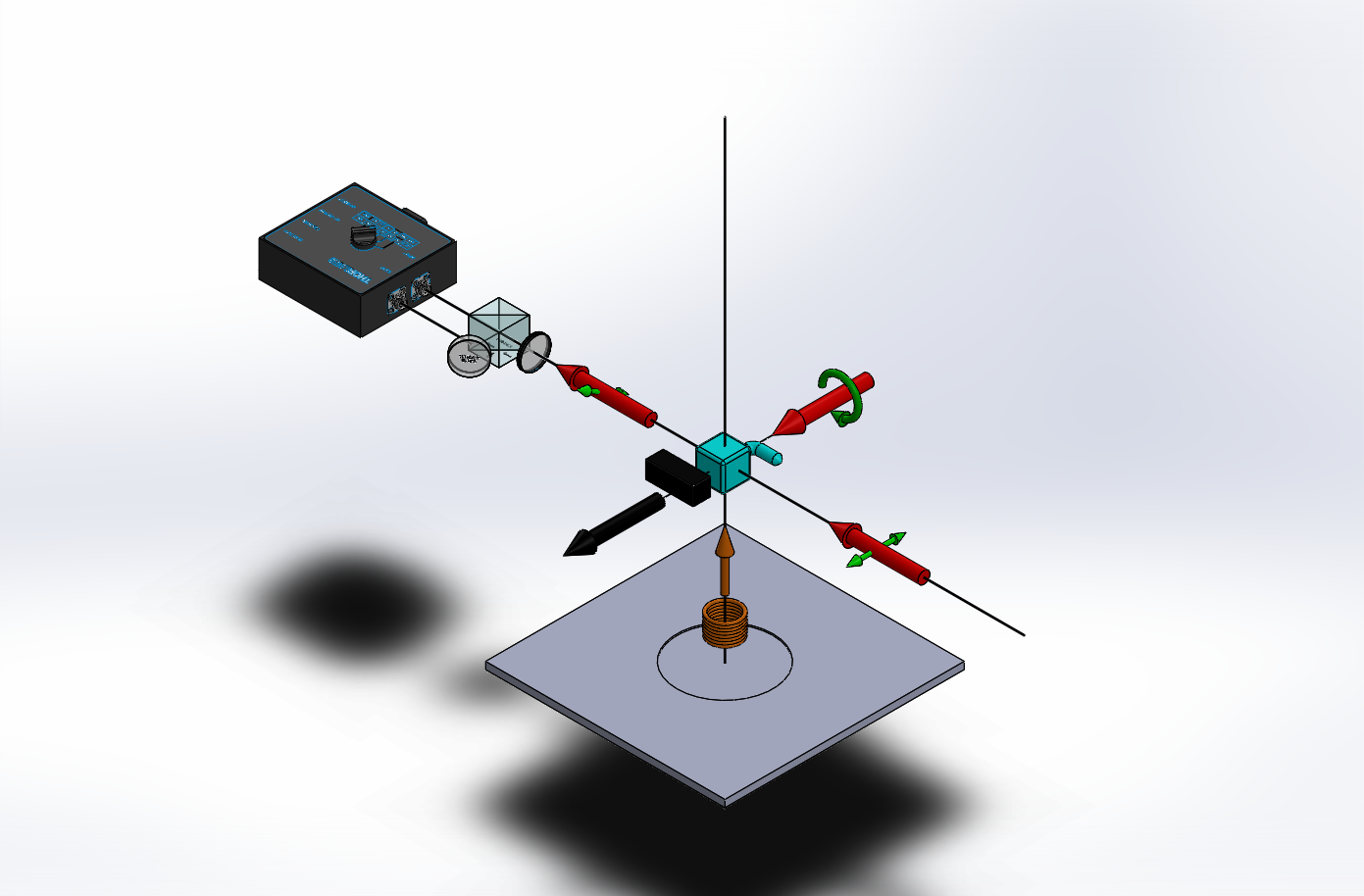} &
         \includegraphics[height=0.45\columnwidth]{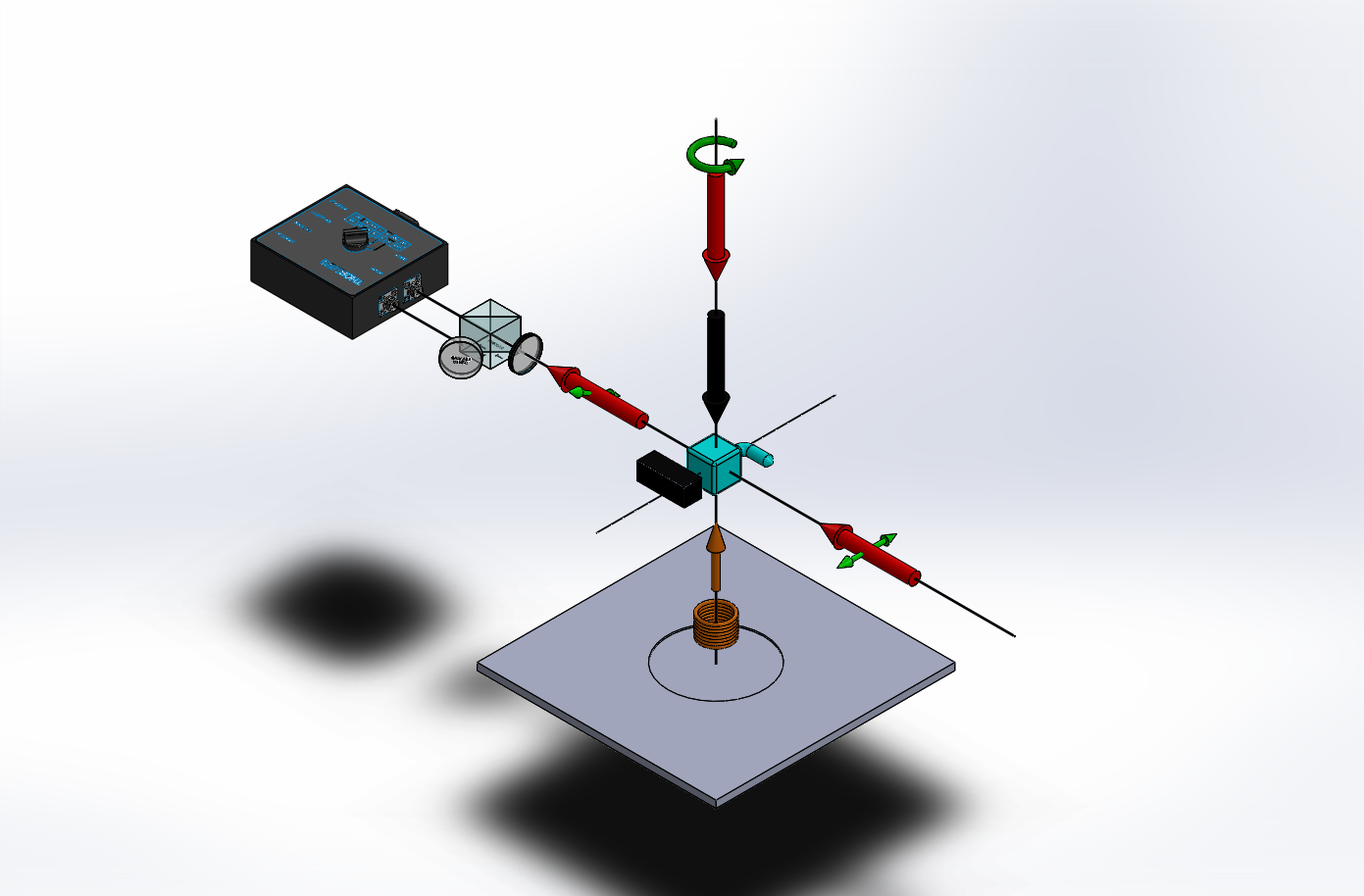} \\
         \small{(a)} & \small{(b)} & \small{(c)} 
    \end{tabular}
\end{center}
\caption{Main components of the experimental setup for (a) the two coil geometry, (b) the primary and (c) secondary configurations. Note: (b) and (c) are referred to in the results section of the paper. The polarisation of the atomic spins along the direction of a static bias magnetic field, represented by the black arrow, is produced by a circularly polarised pump beam. The linearly polarised probe beam monitors the spin precession driven by the rf magnetic field. The polarimetry is performed with a half-wave plate, polarising beam splitter and a balanced photodetector.}
\label{fig:coils}
\end{figure*}

The detection of an rf magnetic field with alkali-metal atomic vapour begins with the optical preparation of the atomic ensemble, which is done via the polarisation of the atomic spins along the direction of a static bias magnetic field $\textbf{B}_{0}$,  Fig.~\ref{fig:coils}(c). This is followed by an interaction with the rf field that is to be measured. Radio-frequency magnetic fields create atomic coherences in the polarised atoms, resulting in the precession of the collective atomic spin components around the bias field. The measurement of the rf field is concluded by an optical readout, i.e., the spin precession is detected through the polarisation rotation of a linearly polarised probe beam via the paramagnetic Faraday effect. All three processes - the preparation, interaction, and readout - are usually performed simultaneously. It is important to point out that the sensor measures rf fields within the plane that is orthogonal to the bias field axis, which is the insensitive axis of the sensor.

In terms of performance, apart from sensitivities at the fT/Hz$^{1/2}$ level \cite{Savukov2005, Chalupczak2012}, atomic magnetometers offer a range of properties that could prove advantageous in various measurement scenarios. 
By monitoring the amplitude and phase of the rf-driven atomic coherence, the magnetometer provides a 2D  vector rf field measurement \cite{Bevington2019b, rushton_2022}. Adjustments to the strength of the bias field tunes the coherence precession frequency and hence facilitates tunabilty of the operational frequency \cite{Savukov2005}. The relatively narrow bandwidth of the sensor, defined by the atomic coherence lifetime, enables filtering out of environmental noise. However, this lifetime typically limits the response rate of the sensor. Feeding back the signal produced by spontaneous fluctuations of the atomic coherences results in a self-oscillating system, often referred to as a spin maser \cite{Bloom1962, Richards1988, Chupp1994, Stoner1996, Bevington2019c, Bevington2019d}. The operation of a magnetometer in the spin maser mode addresses the issue of the sensor's narrow bandwidth. Additionally, the ability to run the measurement as a two-photon process enables operation at low frequencies while maintaining a large static magnetic field, which is therefore less susceptible to perturbations from ambient magnetic field noise \cite{Geng2021, Maddox2023}.

Here, we explore the detection of rf field components with various polarisations by the rf atomic magnetometer. The studies are performed in the context of MIT measurements.
The detection of the polarisation of an rf magnetic field relies on the projection of the measured field onto the polarisation state relevant to that particular sensor.
We begin with considerations regarding the generation and detection of different polarisation states of rf fields. The importance of this discussion is illustrated by numerical modelling via COMSOL simulations and experimental measurements involving different polarisations of the detected rf field. These studies describe the signals measured in scenarios where there are several different inseparable field sources, i.e., the primary field and the secondary field that is generated through eddy currents and magnetisation, as observed in materials such as steel.
Through our modelling and measurements of the inductive signals in MIT measurements, we are able to select the best configuration and strategy for defect or object detection in different materials.

\section{Field measurement}

\subsection{Field polarisation}

\begin{figure}[t]
  \centering
    \includegraphics[width=\linewidth]{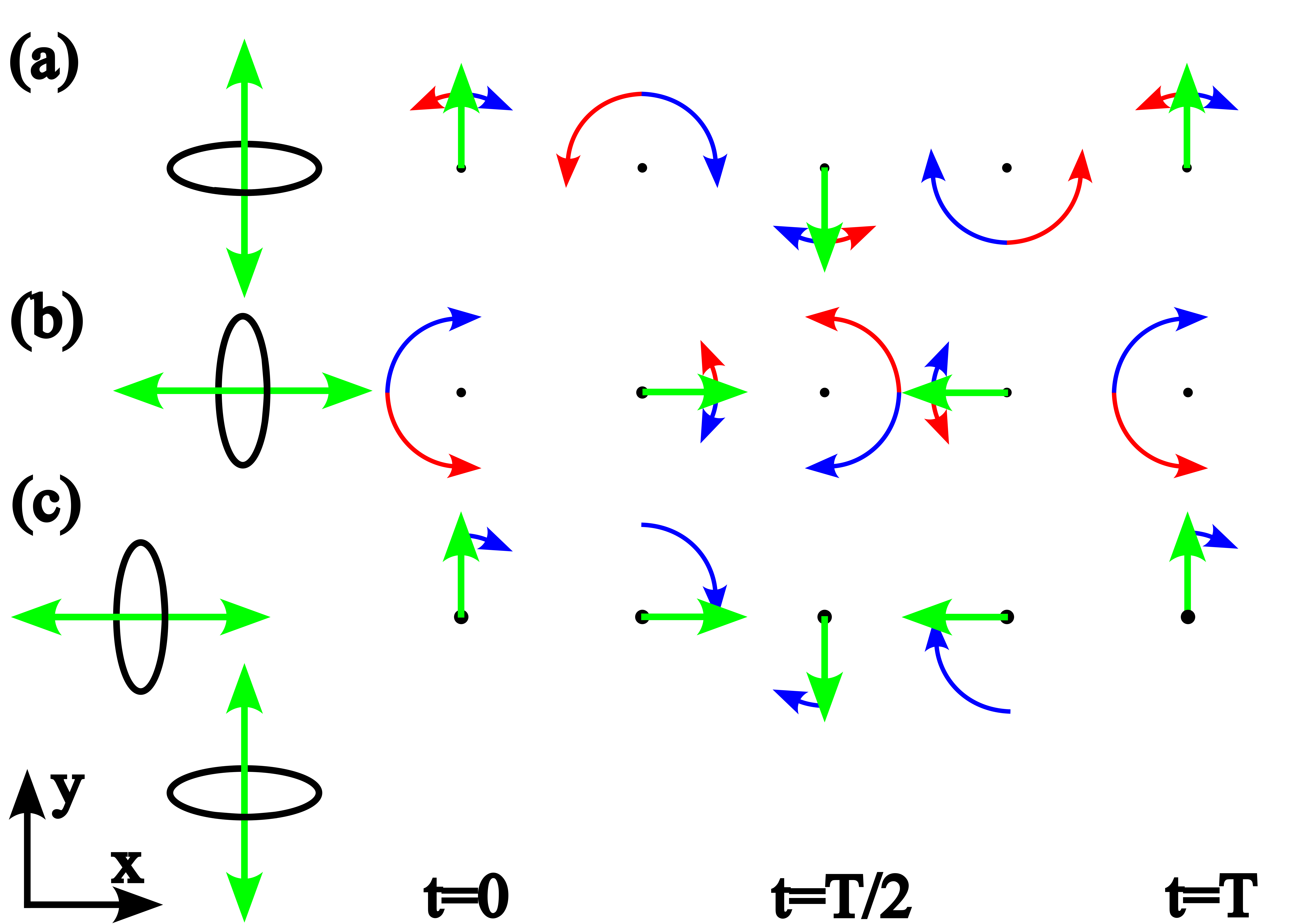}
    \caption{
    Linearly polarised fields (double-ended green arrows) in (a) and (b) are generated by driving an oscillating current through a circular coil, and are equivalent to an equal superposition of two orthogonal circularly polarised rf fields, indicated by red and blue arrows. (c) The sum of two orthogonal linearly polarised fields with a $\Delta \Theta = \phi_{y}-\phi_{x}=90\degree$ phase difference results in the generation of a clockwise rotating field.}\label{fig:Circular_Field}
\end{figure}

Figure~\ref{fig:Circular_Field} presents a simple realization of two linearly, (a) and (b), and one circularly, (c), polarised rf magnetic fields. In particular, Figs.~\ref{fig:Circular_Field}(a) and (b) represent the linearly polarised rf magnetic fields $B_{y}(t)= |B_{y}| \sin(\omega_{\text{rf}} t +\phi_{y}) $ and $B_{x}(t)= |B_{x}| \sin(\omega_{\text{rf}} t +\phi_{x})$, respectively.
This field can be expressed as a superposition of two orthogonal circularly polarised fields, i.e., magnetic fields with constant amplitude and rotating direction, as represented by the red and blue arrows in Figs.~\ref{fig:Circular_Field}(a) and (b). 
Clockwise (CW) circularly polarised rf fields, Fig.~\ref{fig:Circular_Field}(c), can be generated by the combination of $B_{y}(t)$ and $B_{x}(t)$ with $\phi_{y} = 90\degree$, Fig.~\ref{fig:Circular_Field}(a), and $\phi_{x} = 0\degree$, Fig.~\ref{fig:Circular_Field}(b). Counter-clockwise (CCW) circularly polarised rf fields can be generated with $\phi_{y} = 270\degree$ and $\phi_{x} = 0\degree$.

\subsection{Sensor coordinate system}
In the same way that a circular coil is able to produce a linearly polarised field, a pick-up coil sensor would detect an oscillating field by mapping it onto its axis.  
Extension to a total 3D field measurement is achieved by combinations of orthogonal 1D measurements. This sensor category of 1D sensors also includes giant magneto-resistance (GMR) magnetometers \cite{Dogaru2001} and fluxgate detectors \cite{elson_meraki_2022}.

The character of coupling between the atoms and rf field makes the rf atomic magnetometer sensitive to either the CW or CCW circular polarisation component of the rf field, depending on the direction of the bias field and the relevant ground state Land\'e $g$ factor \cite{Georginov2019}. 
Because of this, the rf atomic magnetometer is intrinsically sensitive to orthogonal rf fields, such as $B_{x}(t)$ and $B_{y}(t)$. The magnetometer is not sensitive to oscillating fields along the bias field direction. The sensor's rf field sensitivity is described in detail in the following section.

\subsection{Sensor output}
The output of the atomic magnetometer can be evaluated by solving the Bloch equation for the ensemble of caesium spins with the collective atomic spin $\textbf{J}$ coupled to the sum of bias and rf fields $\textbf{B}=\textbf{B}_{0}+\textbf{B}_{\text{rf}}(t)$, where $\textbf{B}_{\text{rf}}(t) = B_{x}(t)\hat{\textbf{x}} + B_{y}(t)\hat{\textbf{y}}$ \cite{jensen_2019, rushton_2022}. The Bloch equation in the laboratory frame is given by
\begin{equation}
\frac{d\textbf{J}}{dt} = \textbf{J} \times \gamma\textbf{B} + \textbf{J}_{\text{max}} - (R_{p} + \Gamma)\textbf{J},
\label{eq:dJdt_SpinsInStaticandOscillatingMagneticFields}
\end{equation} where
$\gamma$ denotes gyromagnetic ratio, $\textbf{J}_{\text{max}} = R_{p} J_{\text{max}}\hat{\textbf{z}}$ the net momentum of a fully-polarised ensemble, $R_p$ pumping, and $\Gamma$ relaxation rates \cite{graf_2005}. For ease of computation, we rewrite our rf fields in the form $B_{x}(t) = B_{\text{x,c}} \cos (\omega_{\text{rf}} t) + B_{\text{x,s}} \sin (\omega_{\text{rf}} t)$, with $B_{\text{x,c}}=|B_{x}|\cos(\phi_{x})$ and $B_{\text{x,s}}=|B_{x}|\sin(\phi_{x})$ [and equivalent for $B_{y}(t)$]. Equation~\ref{eq:dJdt_SpinsInStaticandOscillatingMagneticFields} is solved by transforming to a new frame, denoted by primes, which rotates about $\hat{\textbf{z}}$ with frequency and direction $\boldsymbol{\omega}_{\text{rf}} = -\omega_{\text{rf}}\ \hat{\textbf{z}}$ \cite{ThorntonBook}: 
\begin{equation}
\frac{d\textbf{J}'}{dt} = \textbf{J}' \times \gamma\left(\textbf{B}' + \frac{\boldsymbol{\omega}_{\text{rf}}}{\gamma}\right)+ \textbf{J}_{\text{max}} - (R_{p} + \Gamma)\textbf{J}'.
\label{eq:dJdt_SpinsInRotFrame}
\end{equation}
The spins $J'_{x}$, $J'_{y}$ and $J'_{z}$ are solved for in the steady state (see supplementary material), i.e. $d\textbf{J}'/dt = 0$, with solutions

\begin{align}
    J'_{x} &= -J_{\text{ss}}\frac{\gamma(\Delta_{\text{rf}}[B_{\text{\text{x,c}}}-B_{\text{y,s}}]  + \delta \omega[B_{\text{x,s}}+B_{\text{y,c}}] )/2}{\delta \omega^{2} + \gamma^{2}  (\langle B_x' \rangle^2 + \langle B_y' \rangle^2) + \Delta_{\text{rf}}^{2}},
\label{eq:Jx'}\\
J'_{y} &= J_{\text{ss}}\frac{\gamma (\delta \omega [B_{\text{\text{x,c}}}-B_{\text{y,s}}]- \Delta_{\text{rf}}[B_{\text{x,s}}+B_{\text{y,c}}] ) /2}{\delta \omega^{2} + \gamma^{2} (\langle B_x' \rangle^2 + \langle B_y' \rangle^2) + \Delta_{\text{rf}}^{2}},
\label{eq:Jy'}\\
J'_{z} &= J_{\text{ss}} \frac{\delta \omega^{2} + \Delta_{\text{rf}}^{2}}{\delta \omega^{2} + \gamma^{2} (\langle B_x' \rangle^2 + \langle B_y' \rangle^2) + \Delta_{\text{rf}}^{2}},
\label{eq:Jz'}
\end{align}
where $J_{\text{ss}} = R_p J_{\text{max}}/\delta\omega$, $\delta\omega = R_{p} + \Gamma$  and $\Delta_{\text{rf}} = \omega_\text{rf}-\gamma B_0$. Demodulating this signal, we identify the in- and out-of-phase components as $X\propto J_{y}'$ and $Y\propto J_{x}'$. Moreover, when $\omega_{\text{rf}}=\omega_{L}$ ($\Delta_{\text{rf}}=0$) and in the limit $\delta\omega^2 \ll \gamma^2(\langle B_x' \rangle^2 + \langle B_y' \rangle^2)$ - both of which are always satisfied in this paper - we find \cite{rushton_2022}
\begin{equation}
    X\propto B_{\text{x,c}}-B_{\text{y,s}},
    \label{eq:LockinX}
\end{equation}
\begin{equation}
   Y\propto -(B_{\text{x,s}}+B_{\text{y,c}}). 
   \label{eq:LockinY}
\end{equation}
From $X$ and $Y$, the signal amplitude and signal phase of the rf atomic magnetometer can be calculated as
\begin{equation}
    \text{Signal Amp.} = \sqrt{X^{2}+Y^{2}}
    \label{eq:SignalAmp}
\end{equation}
and
\begin{equation}
    \text{Signal Phase} = \arctan(\frac{Y}{X}),
\label{eq:SignalPhase}
\end{equation}
respectively. This result shows the fundamental differences in the behaviour of rf atomic magnetometers to other 1D-linear sensors, as the $B_{x}(t)$ and $B_{y}(t)$ rf fields can ``mix'' in Eqs.~\ref{eq:LockinX} and \ref{eq:LockinY}.


\begin{figure}
 \centering
    \begin{tabular}{cc}
        \includegraphics[width=0.5\linewidth]{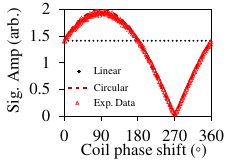} &  \includegraphics[width=0.5\linewidth]{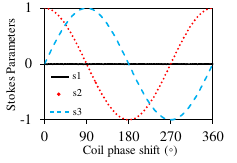}\\
         a) & c) \\
          \includegraphics[width=0.5\linewidth]{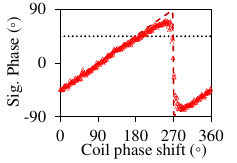} &  \includegraphics[width=0.4\linewidth]{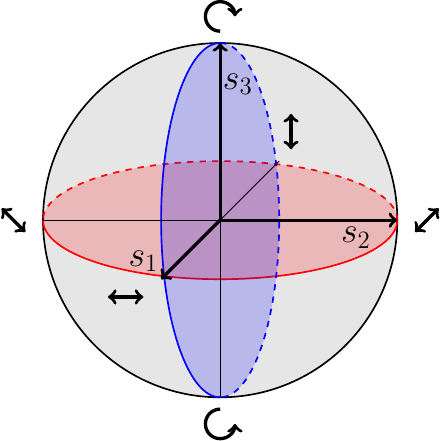}\\
        b) & d) \\
    \end{tabular}
    \caption{(a) Amplitude and (b) phase of the magnetometer signal monitoring the field $\textbf{B}_{\text{rf}}(t) = B_{x}(t)\hat{\textbf{x}} + B_{y}(t)\hat{\textbf{y}}$, as shown in Fig.~\ref{fig:Circular_Field}(c), as a function of the phase difference between fields. The amplitude and phase are simulated in COMSOL for a linear 1D detector (dotted black line) and for a circular detector (dashed red line), i.e., an rf magnetometer. Red triangles represent measurements with an rf atomic magnetometer. (c) Stokes parameters calculated for the combined field, $\textbf{B}_{\text{rf}}(t)$, from simulated COMSOL data. These parameters trace out the polar plane outlined by the solid black line in the Poincar\'e sphere, (d).}
    \label{fig:VectorLockin}
\end{figure}

To illustrate the differences between an rf atomic magnetometer and 1D sensors we consider the measurements of the rf field produced by two identical coils, one directed along the $x$-axis and the other along the $y$-axis, as depicted in Figs.~\ref{fig:coils}(a) and \ref{fig:Circular_Field}(c).
The total magnetic field is measured at the intersection of the axes of the two coils, which is at an equal distance from both coils.

Figure~\ref{fig:VectorLockin} shows the amplitude, (a), and phase, (b), of the signal as a function of the relative phase, $\Delta\Theta$ ($0\degree$-$360\degree$), between $B_{x}(t)$ and $B_{y}(t)$. The change of the relative phase results in a change of the field polarisation from linear at $\Delta\Theta = 0\degree$ and $180\degree$ to circular at $\Delta\Theta = 90\degree$ and $270\degree$. 
The dotted and dashed lines in Fig.~\ref{fig:VectorLockin} represent data simulated in a 2D ($x$- and $y$-axes) COMSOL model. The calculation of the resultant field amplitude and phase are performed in two different ways. The first (dotted black line) assumes $B_{x}$ and $B_{y}$ are measured by two independent linear 1D sensors, for example, pick-up coils. Combining the two fields in a vector measurement would result in a total field with an amplitude $|B_{T}| = \sqrt{|B_{y}|^2+|B_{x}|^2}$ and direction $\theta_{B_{T}} = \text{tan}^{-1}({|B_{y}|/|B_{x}|)}$, which is represented by the black dotted line in Figs.~\ref{fig:VectorLockin}(a) and (b), respectively. The output of the 1D sensor does not show any amplitude or phase variation. 
The red dashed lines in all figures are the results of COMSOL calculations modelling a circular (phase-sensitive) 2D sensor, i.e., an rf atomic magnetometer, which is governed by Eqs.~\ref{eq:LockinX}-\ref{eq:SignalPhase}. 
The red triangles in Figs.~\ref{fig:VectorLockin}(a) and (b) are data recorded with the rf atomic magnetometer. The maximum amplitude at $\Delta\Theta=90\degree$ and the minimum amplitude at $\Delta\Theta=270\degree$ correspond to the measurements of orthogonal circularly polarised fields \cite{Georginov2019, motamedi2023}. 

Figure~\ref{fig:VectorLockin}(c) describes the polarisation of the rf field, produced by two coils, in terms of the Stokes parameters $S_{0,1,2,3}$, normalised such that $s_1^2 + s_2^2 + s_3^2 = 1$, where $s_{1,2,3} = S_{1,2,3}/S_{0}$. These define the total field's energy and projections onto vertical/horizontal linear, $\pm45\degree$ linear and left/right-handed circular polarisation axes, respectively. These definitions (in terms of initial rf fields) and their relation to the Poincar\'e sphere, Fig.~\ref{fig:VectorLockin}(d), are made explicit in \cite{Walkenhorst2020}, as is a comment on the difficulty of representing polarisation evolution on a sphere in a publication or display. As the coils sweep through a phase difference of $360\degree$ in the data in Fig.~\ref{fig:VectorLockin}, we see the total field cycles between perfect $\pm45\degree$ linear polarisations (at $\Delta\Theta = 0\degree, 180\degree$ and $360\degree$), and perfect left and right-handed circular polarisations (at $\Delta\Theta = 90\degree$ and $270\degree$, respectively). The normalised Stokes parameters plotted on the Poincar\'e sphere would trace out the polar plane outlined by the solid black line in Fig.~\ref{fig:VectorLockin}(d), and is graphically represented in a movie showing the polarisation evolution in the supplementary material. It is worth pointing out that the signal amplitude can be expressed through rf field Stokes parameters $R = \sqrt{S_{0} - S_{3}}$, which explicitly confirms the dependence of the sensor output on the polarisation. The consequences of this dependence will be discussed in the context of MIT signals modelled for a composite material which has both significant electrical conductivity and magnetic permeability, such as stainless steel.

\section{Secondary Field Sources}

Considerations from the previous section become important for rf atomic magnetometer-based MIT measurements, where contributions from the different field sources, i.e., primary and secondary fields, result in changes not only in the signal amplitude and phase, but also polarisation of the detected field. 

Penetration of the rf primary magnetic field into the object has a dissipative character with an attenuation typically parameterised by the skin depth. Previous work has reported on the different physical characteristics of secondary field generation due to a material's properties being dominated by either electrical conductivity or magnetic permeability \cite{Gartman2021}. 
In highly electrically conductive objects, e.g., copper with $\sigma \sim 60 ~\text{MS}/\text{m}$, a primary rf field induces eddy currents that generate an rf secondary field with a phase-shift with respect to the driving field equal to 180$\degree$ at high frequencies, and 90$\degree$ at low frequencies \cite{Bidinosti2007, Honke2018, elson_meraki_2022, rushton_2022}. The eddy current density, also called the conduction current density or free current density, increases and is more confined to the materials' surface with increasing rf frequency.

In materials with a high magnetic permeability, e.g., ferrite with $\mu_{r} \sim 2000$, we see the secondary field dominated by the magnetic moment of bound electrons. In particular, the bound electron's orbit about their nucleus traces a small current loop that is perceived macroscopically as a dipole. Under some external field, these dipoles feel a torque which orients them along said field's axis, generating a measurable, net magnetisation \cite{GriffithsBook}.

Each of these contributions to the secondary field are made explicit in the following form of Ampere's law, where $\textbf{j}_m$ defines a magnetisation current density and $\textbf{j}_c$ defines the previously mentioned conduction density. These give rise to the expected  magnetisation and eddy-current generated fields, respectively \cite{CoeyBook}
\begin{equation}
\frac{1}{\mu_0}[\nabla \cross \textbf{B}_{\text{rf}}(t)] = \textbf{j}_{c} + \textbf{j}_{m}.
\label{Ampere}
\end{equation}

Here, $\textbf{B}_{\text{rf}}(t)$ describes the total measured field, i.e. both the primary and secondary responses. We note that Eq.~\ref{Ampere} describes linear, isotropic media. As such, hysteretic effects from ferromagnetic materials are not included in this model. 

In certain geometries, we may derive an analytical expression for $\textbf{B}_{\text{rf}}(t)$ and identify its dependence on some induced magnetic moment $\textbf{m}$: Bidinosti et al. finds such an expression for a magnetically-permeable, conducting sphere in a uniform ac magnetic field \cite{Bidinosti2007}. We may then identify $\textbf{m}$ for limiting cases of highly conductive or magnetic materials, at similar extremes of frequency.
 
\begin{figure}[t]
    \centering
    \includegraphics[width=\linewidth]{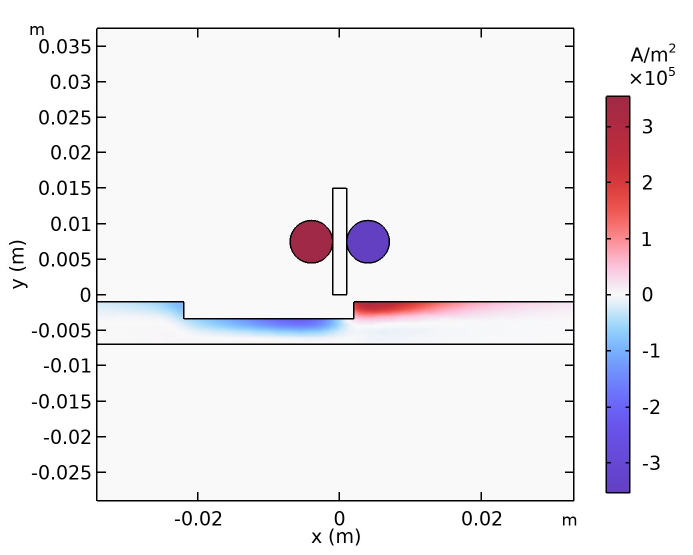}
    \caption{COMSOL model. The $z$-component of the eddy current density $j_{\text{c,z}}$ is plotted inside a plate with $\sigma=3$~MS/m and $\mu_{r}=1$, as well as the current density in the excitation coil. The magnetic field measured at the vapour cell $(x=0, y=0.1~\text{m})$ is integrated over its surface in the COMSOL simulations. The centre of the plate is scanned from $x=-150$~mm to $x=150$~mm in 1~mm increments for each line scan. The coil consists of a 2~mm diameter ferrite core, surrounded by a copper coil.}\label{fig:COMSOLModel}
\end{figure}

We can identify defects in an object where the direction of the secondary field is affected by its geometry. Recalling their generation from free electrons: eddy currents will concentrate near boundaries, i.e., the edges of cracks, Fig.~\ref{fig:COMSOLModel}, resulting in a non-uniform current distribution which generates field components orthogonal to the primary field. Conversely, the size and direction of the magnetisation field has volume character: as all the atomic dipoles in the materials are brought into alignment by the primary field, they form a large pseudo-dipole, whose net magnetic field diverges around the edges of the object.

\begin{figure*}
     \centering
     \begin{subfigure}[b]{0.329\textwidth}
         \centering
         \includegraphics[width=\textwidth]{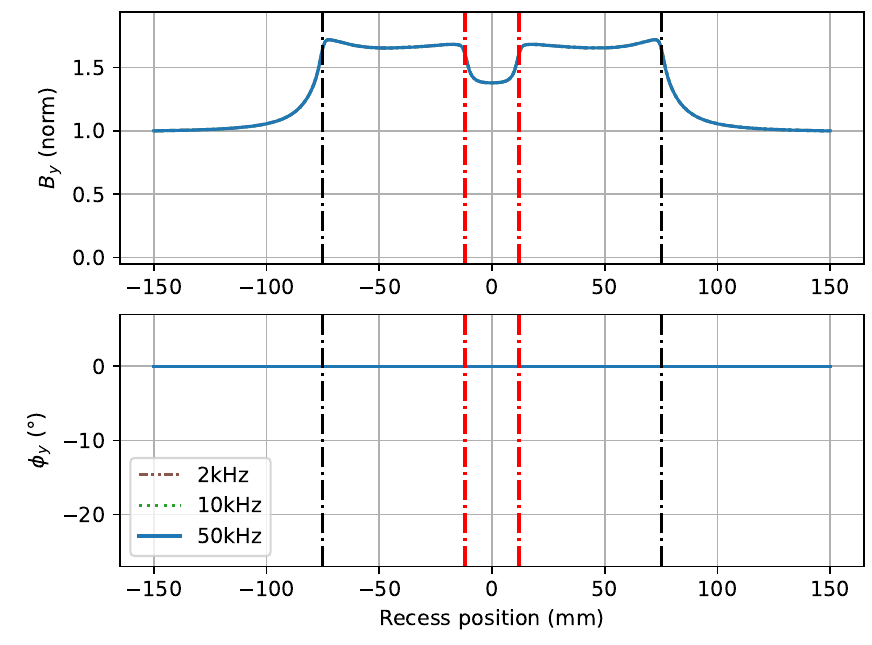}
         \caption{$B_{y}$}
     \end{subfigure}
     \hfill
     \begin{subfigure}[b]{0.329\textwidth}
         \centering
         \includegraphics[width=\textwidth]{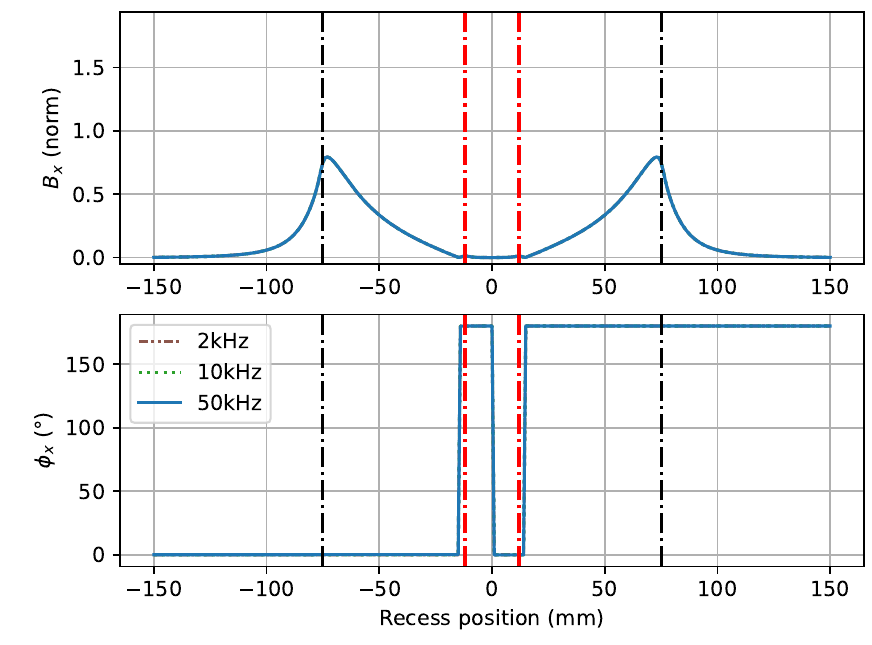}
         \caption{$B_{x}$}
     \end{subfigure}
     \hfill
     \begin{subfigure}[b]{0.329\textwidth}
         \centering
         \includegraphics[width=\textwidth]{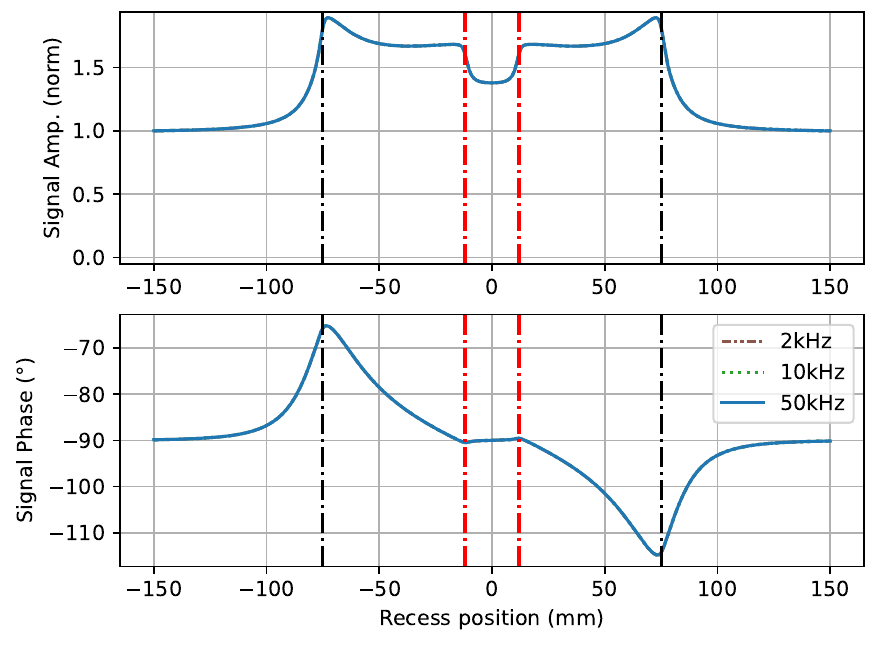}
         \caption{Signal amplitude and phase}
     \end{subfigure}
     \begin{subfigure}[b]{0.329\textwidth}
         \centering
         \includegraphics[width=\textwidth]{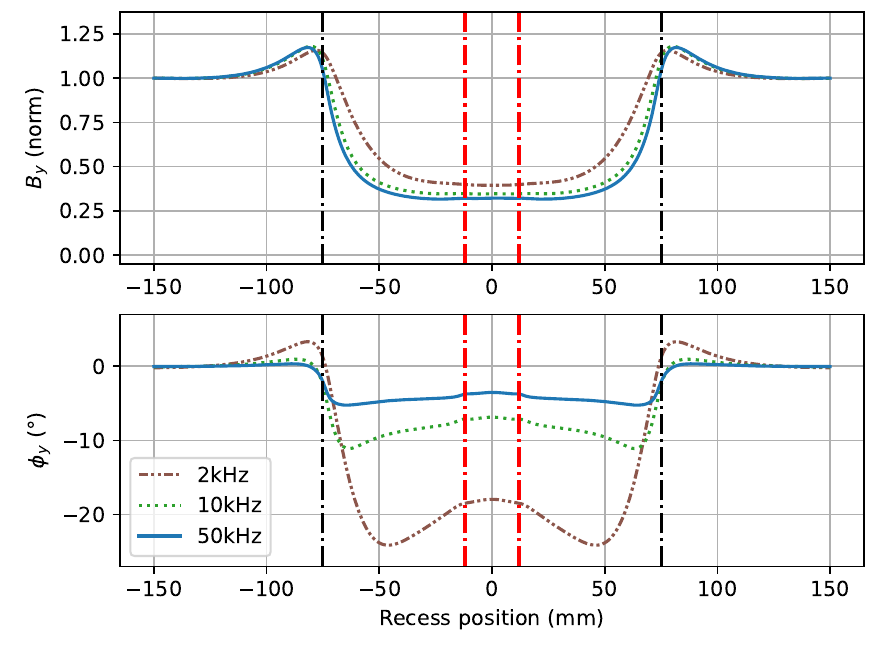}
         \caption{}
     \end{subfigure}
     \hfill
     \begin{subfigure}[b]{0.329\textwidth}
         \centering
         \includegraphics[width=\textwidth]{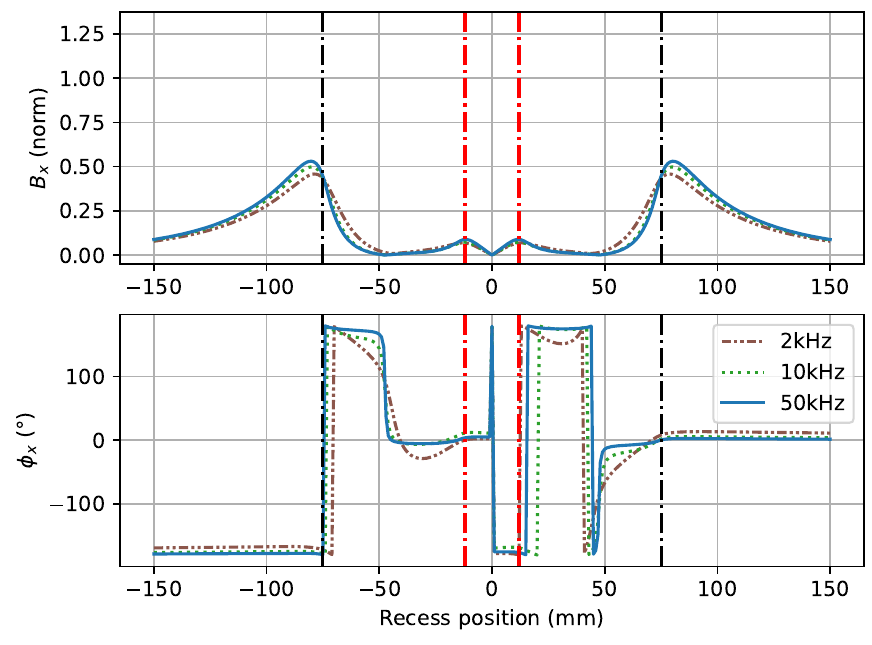}
         \caption{}
     \end{subfigure}
     \hfill
     \begin{subfigure}[b]{0.329\textwidth}
         \centering
         \includegraphics[width=\textwidth]{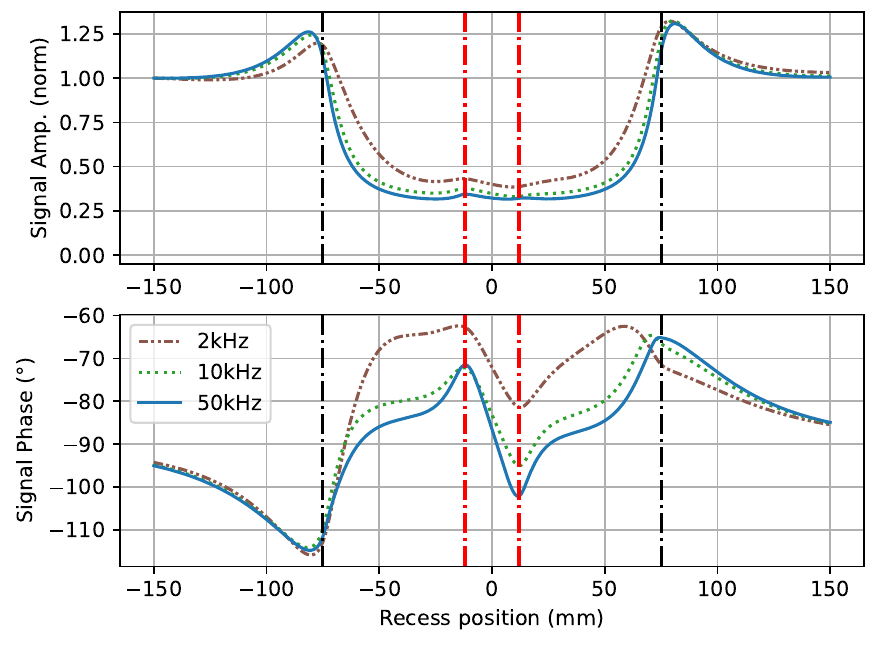}
         \caption{}
     \end{subfigure}
     \begin{subfigure}[b]{0.329\textwidth}
         \centering
         \includegraphics[width=\textwidth]{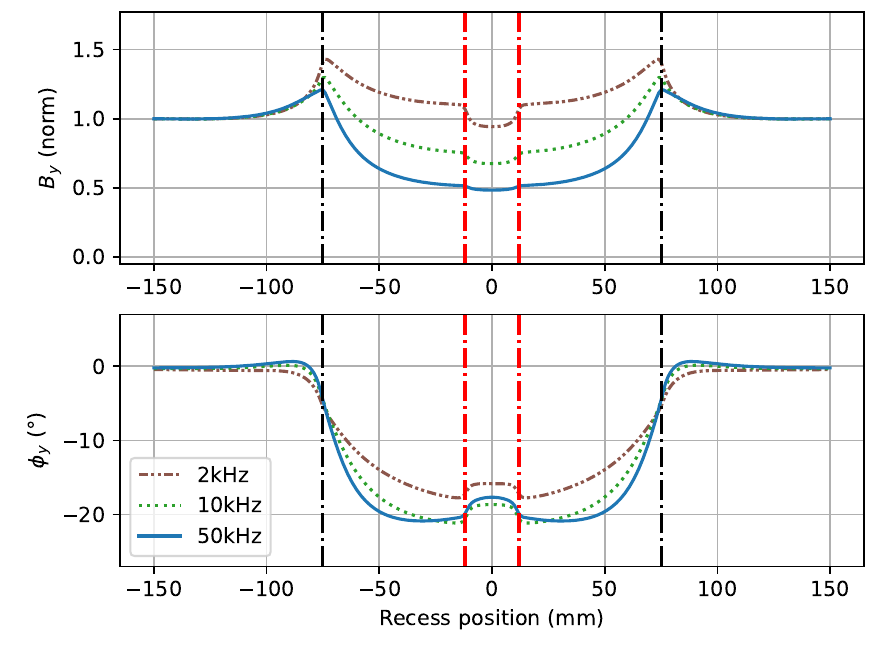}
         \caption{}
     \end{subfigure}
     \hfill
     \begin{subfigure}[b]{0.329\textwidth}
         \centering
         \includegraphics[width=\textwidth]{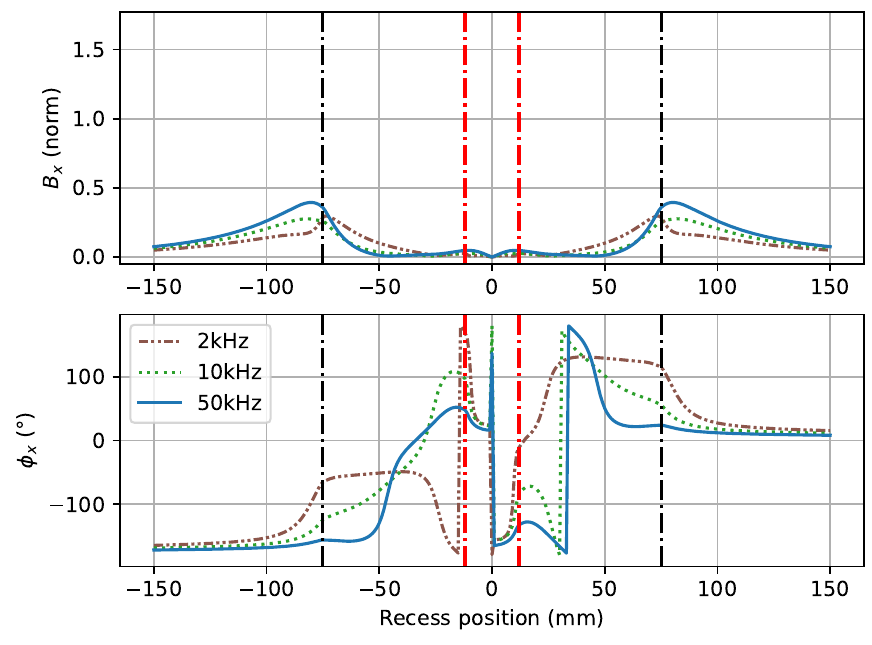}
         \caption{}
     \end{subfigure}
     \hfill
     \begin{subfigure}[b]{0.329\textwidth}
         \centering
         \includegraphics[width=\textwidth]{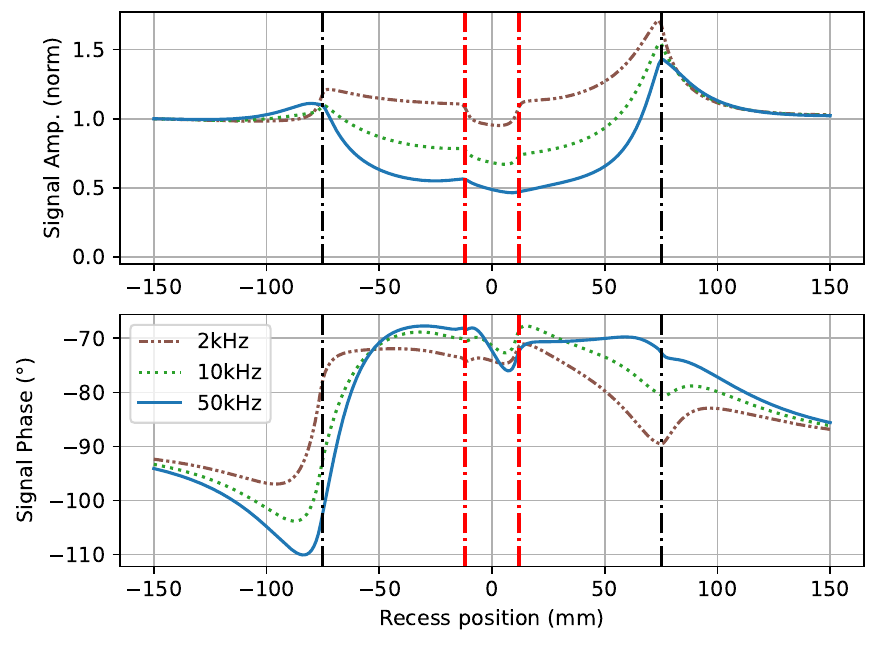}
         \caption{}
     \end{subfigure}
        \caption{COMSOL data of the amplitude and phase (top and bottom row in each figure) of (a, d, g) $B_{y}$, (b, e, h) $B_{x}$, and (c, f, i) the signal amplitude and phase (Eqs.~\ref{eq:SignalAmp}-\ref{eq:SignalPhase}) for an rf atomic magnetometer, recorded while scanning over a $\SI{150}{\milli\meter} \times \SI{6}{\milli\meter}$ plate with a $\SI{24}{\milli\meter} \times \SI{2.4}{\milli\meter}$ recces in its center. The plate is modelled as (a)-(c) $\mu_{r}=80$ and $\sigma=\SI{1}{\milli\siemens\per\meter}$, (d)-(f) $\mu_{r}=1$ and $\sigma=\SI{3}{\mega\siemens\per\meter}$, and (g)-(i) $\mu_{r}=80$ and $\sigma=\SI{3}{\mega\siemens\per\meter}$. The edges of the plate and the recess are denoted by black and red dotted lines, respectively.
        }
        \label{fig:COMSOL_Magnetic_Conductor_Mix}
\end{figure*}

\subsection{Model} 
\label{sec:COMSOLSection}
COMSOL is used to model signals in an MIT measurement, where an object with a defined electrical conductivity $\sigma$ and magnetic permeability $\mu=\mu_{r}\mu_{0}$ is scanned under the rf primary field coil, Fig.~\ref{fig:COMSOLModel}, and the primary and secondary rf magnetic fields are measured in a sample volume. We consider a square plate, with $\SI{150}{\milli\meter}$ length and $\SI{6}{\milli\meter}$ thickness, containing a $\SI{24}{\milli\meter}$ diameter recess that is $\SI{2.4}{\milli\meter}$ deep. The scan is performed across the centre of the plate.

Figure~\ref{fig:COMSOL_Magnetic_Conductor_Mix} shows the amplitude and phase of the rf fields $B_{y}$, (a), and $B_{x}$, (b), for a ferrite-like plate with the parameters $\mu_{r}=80$ and $\sigma\sim0$, calculated for three different frequencies of $\SI{2}{\kilo\hertz}$, $\SI{10}{\kilo\hertz}$ and $\SI{50}{\kilo\hertz}$ (that lie on top of each other in this scale).
Based off Eqs.~\ref{eq:LockinX}-\ref{eq:SignalPhase}, the amplitude and phase of the signal detected by the rf atomic magnetometer are plotted in Fig.~\ref{fig:COMSOL_Magnetic_Conductor_Mix}(c). The same analysis is done for an electrically conductive plate, with the parameters $\mu_{r}=1$ and $\sigma=3$~MS/m, Figs.~\ref{fig:COMSOL_Magnetic_Conductor_Mix}(d-f), as well as for a composite material, representing a carbon steel material with the parameters $\mu_{r}=80$ and $\sigma=3$~MS/m, Figs.~\ref{fig:COMSOL_Magnetic_Conductor_Mix}(g-i). We note that carbon steel is ferromagnetic, however we only explore its permeable and conductive properties in COMSOL. 

The magnetically permeable and conductive objects represent orthogonal materials, whose secondary field responses are dominated by magnetisation and eddy current effects, respectively. There are signatures of the recess and plate edges visible in each plot. Considering the signatures of the recess feature (marked with red dot-dashed line) in the detected signal amplitude in  Figs.~\ref{fig:COMSOL_Magnetic_Conductor_Mix}(c) and (f), it can be seen that the greatest amplitude change mirrors $B_{y}$ for ferrite-like plates, Fig.~\ref{fig:COMSOL_Magnetic_Conductor_Mix}(a), and $B_{x}$ for the electrically conductive object, Fig.~\ref{fig:COMSOL_Magnetic_Conductor_Mix}(e). 
The phase information for $B_{x}$ and $B_{y}$ is more complicated to analyse, partially due to the changing direction of the $B_{x}$ component across the plate. The phase data of $B_{x}$ and $B_{y}$ is symmetric, however wrapping/$x=0$ inversion means that they do not appear so here. Comparing the results in Figs.~\ref{fig:COMSOL_Magnetic_Conductor_Mix}(c) and (f), the recess signature is more pronounced in the amplitude data in Fig.~\ref{fig:COMSOL_Magnetic_Conductor_Mix}(c) and in the phase data in Fig.~\ref{fig:COMSOL_Magnetic_Conductor_Mix}(f). The data in Fig.~\ref{fig:COMSOL_Magnetic_Conductor_Mix}(f) is slightly asymmetric, due to the fact that the sample is not infinitely conductive.
Figures~\ref{fig:COMSOL_Magnetic_Conductor_Mix}(g-i), which represent results of a simulation for a composite material that is meant to mimic a carbon steel plate, show that the  amplitude and phase of the secondary field features $B_{x}$ and $B_{y}$ produced by the plate edges are symmetric, while relevant features in the detected signal amplitude and phase are asymmetric. This is due to the change of the polarisation of the rf field at the sensor location.
The expectation for the steel plate data is that its amplitude response would be more similar to a magnetically permeable object at low frequencies, Fig.~\ref{fig:COMSOL_Magnetic_Conductor_Mix}(b), and an electrically conductive object at high frequencies, Fig.~\ref{fig:COMSOL_Magnetic_Conductor_Mix}(e) \cite{Gartman2021}. This is explored in more detail in Sec.~3 of the supplementary material using theory from \cite{Bidinosti2007}. For a magnetically permeable plate, the edge signal can be characterised by the steep decay of the edge signal outside of the plate and a shallow decay inside, while the opposite is true for electrically conductive objects, and is visible in the data in Fig.~\ref{fig:COMSOL_Magnetic_Conductor_Mix}(h).

The polarisation of the rf field seen by the sensor changes over the objects with both non-negligible magnetic permeabilities and electrical conductivities, e.g., the composite material representing a carbon steel sample. To analyse this in detail, Stokes parameters of the field were calculated based on previously modelled values at 2~kHz from Figs.~\ref{fig:COMSOL_Magnetic_Conductor_Mix}(g) and (h). 

\begin{figure}[t]
    \centering
    \includegraphics[width=\linewidth]{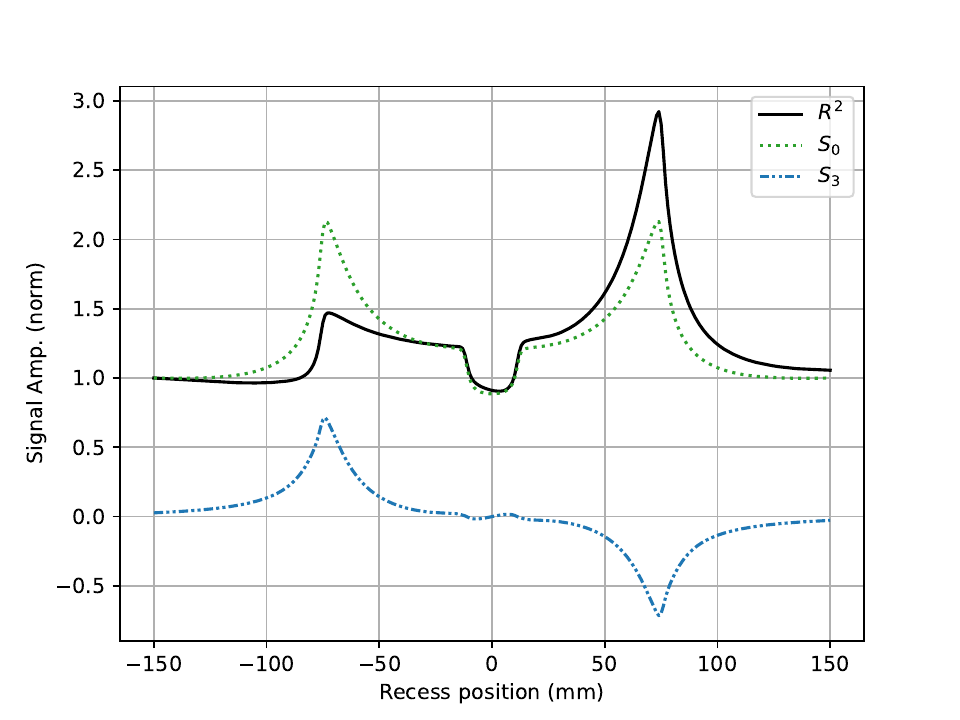}
    \caption{Stokes vector representation. The 2~kHz data from Fig.~\ref{fig:COMSOL_Magnetic_Conductor_Mix}(i) is squared and plotted, alongside the Stokes parameters $S_{0}$ and $S_{3}$, which are calculated from the 2~kHz data in Figs.~\ref{fig:COMSOL_Magnetic_Conductor_Mix}(g) and (h).}\label{fig:S0S3R}
\end{figure}

Figure~\ref{fig:S0S3R} shows the line scan of $S_{0}$ and $S_{3}$ across the plate, which represent the amplitude and circular polarisation of the field, respectively (defined as $B_{x}^{2}+B_{y}^{2}$ and $2B_{x}B_{y}\sin(\phi_{y}-\phi_{x})$ in the supplementary material). The data in Fig.~\ref{fig:S0S3R} has been scaled by the value of $S_0$ at $\pm$ 150~mm. 

The change in $S_{3}$ between 0.7 and -0.7 indicates the change between CCW and CW polarisations, respectively, which affects the sensor's ability to detect the field. In the case that $s_{3}=+1$ then all the rf field will be CCW and no coherences will be generated within the Cs atoms, as we are only sensitive to CW oscillating magnetic fields, in this example. It is worth pointing out that the signal in Fig.~\ref{fig:S0S3R} is always non-zero because this condition is never met.

The $S_{3}$ Stokes parameters are at least three orders of magnitude smaller than $S_{0}$ for the objects whose secondary fields are dominated by either magnetic permeability or electrical conductivity.

In the measurements of $B_{y}$, a significant component of the monitored field is from the primary magnetic field, which accounts for the offset in the background measurement away from the plate. Generally this reduces the overall contrast of the measurement, and can hide small signals, for example those from small defects. This can be cancelled, typically with a coil that is far from the plate and close to the vapour cell \cite{Bevington2019b, jensen_2019, deans_2020, rushton_2022}, or by aligning the sensor's insensitive axis parallel to $B_{y}$ and only monitoring the $B_{x}$ component.

Brief analysis, presented above, of the secondary fields and detected signals, based on results shown in Fig.~\ref{fig:COMSOL_Magnetic_Conductor_Mix}, indicates that it can be beneficial to monitor different secondary field components, e.g., $B_{x}$ and $B_{y}$ or $B_{x}$ and $B_{z}$, for materials with different magnetic properties, and at different frequencies depending upon the penetration depth required.

\section{Experimental Setup}
\label{sec:ExpSetup}
\subsection{rf atomic magnetometer}

The rf atomic magnetometer is operated in a magnetically unshielded environment. A detailed description of the sensor and instrumentation is presented elsewhere \cite{Bevington2018, Bevington2019, Bevington2019b}. Here we limit the discussion to its major components. Atomic magnetometer instrumentation includes four major subsystems: the caesium atomic vapour cell, the magnetic field control system, the lasers, and the detection system. 
The vapour cell has a paraffin anti-relaxation coating to minimise atomic depolarisation via wall collisions. The cell with dimensions  $(10\times10\times10)$~mm$^{3}$ is kept at ambient temperature, atomic density $n_{\text{Cs}}=0.33 \times10^{11}$~$\text{cm}^{-3}$, in a static magnetic bias field $\textbf{B}_{0}$, created by a set of three nested orthogonal square Helmholtz coils, with side lengths $\SI{1000}{mm}$, $\SI{940}{mm}$ and $\SI{860}{mm}$. The magnitude $B_{0}$ defines the rf resonance (Larmor) frequency, i.e., the operating frequency of the sensor. The field is stabilised using a feedback control loop consisting of a three-axis Bartington Mag960 fluxgate (error signal) close to the vapour cell and an SRS960 PID controller (control signal), whose output modulates the current supplied to the square Helmholtz coils (feedback). 
The laser system produces two beams, the pump and the probe beam. The pump beam is circularly polarised and has its frequency stabilized to the $6\,^2$S$_{1/2}$ $F=3\rightarrow{}6\,^2$P$_{3/2}$ $F'=2$ resonance transition (D2 line, $\SI{852}{\nano\meter}$). It propagates parallel to $\textbf{B}_{0}$ and creates a population imbalance within the atomic energy levels of the ensemble of caesium atoms. The probe laser's frequency is red detuned by $\SI{2.75}{\giga\hertz}$ from the $6\,^2$S$_{1/2}$ $F=3$ $\rightarrow{}6\,^2$P$_{3/2}$ $F'=2$ resonance transition and propagates orthogonally to the pump beam. Atomic coherences created by the coupling of the atoms and an rf magnetic field resonant with the Larmor frequency are mapped onto the probe beam's polarisation. The detection is done by monitoring the probe beam's polarisation rotation with a polarimeter, formed by a polarising beamsplitter and a balanced photodiode. The photodiode signal is demodulated by a lock-in amplifier referenced to the primary rf field frequency. 

\subsection{MIT measurement}
The MIT measurement is performed by scanning a target object under the primary rf field coil, which is co-located directly under the cell on the $y$-axis, Fig.~\ref{fig:coils}. The object signature is the relative change in the amplitude and phase of the signal recorded by the lock-in, Eqs.~\ref{eq:SignalAmp} and \ref{eq:SignalPhase}, as the object is scanned under the primary coil. 
The primary coil is located 100~mm from the cell (stand-off) and $\sim$1~mm above the object (lift-off). 
The primary coil has 100 turns and inner-, outer-diameter and length dimensions of 2~mm, 4~mm and 4~mm, respectively, and was wound around the middle of a 2~mm diameter ferrite core that is 7~mm long. The lift-off is measured from the bottom of the ferrite core. 
The object position is moved relative to the primary coil in the $x-z$ plane using two pairs of orthogonal stepper motors. The sample is supported by a plastic frame that is coupled and supported by high-tensile rods connected to the stepper motor platforms, so that the sample is supported on all sides. The stepper motors used can achieve sub-$\mu$m resolution, though larger translational increments of $\sim$2mm are used in this study. The movement time is negligible relative to the data acquisition time to record a full rf resonance (minimum 1~s) that is required per data point.

\section{SENSING GEOMETRY}

By changing the orientation of the bias magnetic field, $\textbf{B}_{0}$, the sensing plane of the rf magnetometer can be changed to: (i) measure one component, $B_{x}$, which is parallel to the surface of the plate and the other, $B_{y}$, orthogonal to the surface of the plate with $\textbf{B}_{0}=B_{0}\hat{\textbf{z}}$, or (ii) only monitor the components $B_{x}$ and $B_{z}$ parallel to the surface of the plate with $\textbf{B}_{0}=B_{0}\hat{\textbf{y}}$. These arrangements are referred to as the primary, and the secondary (or self-compensation in previous work) configuration, respectively. In the latter case, only components of the secondary magnetic field are detectable, while the former will also measure the primary magnetic field directed along the $y$-axis. Measurements were carried out for a carbon steel plate and an aluminium plate in these two sensing configurations and the data are shown in Figs.~\ref{fig:PrimarySecondary}(a-b) and Figs.~\ref{fig:PrimarySecondary}(c-d), respectively. 
Each pixel represents the amplitude and phase (top and bottom row in each figure) of the rf resonance recorded at each position across the plate. All of the amplitude data is normalised by the data points recorded far from the aluminium plate ($\pm 65$~mm) in the primary configuration, Fig.~\ref{fig:PrimarySecondary}(b), which calibrates any frequency dependence of the rf field generated by the coil. Additionally, the amplitude data is multiplied by the rf resonance linewidth for each pixel to account for any amplitude change due to any broadening due to magnetic field gradients, which becomes an issue at low bias field strengths for steel.

\section{Results}
\subsection{Primary configuration}
\begin{figure*}
    \centering
     \begin{subfigure}[b]{0.49\textwidth}
         \centering
         \includegraphics[width=\textwidth]{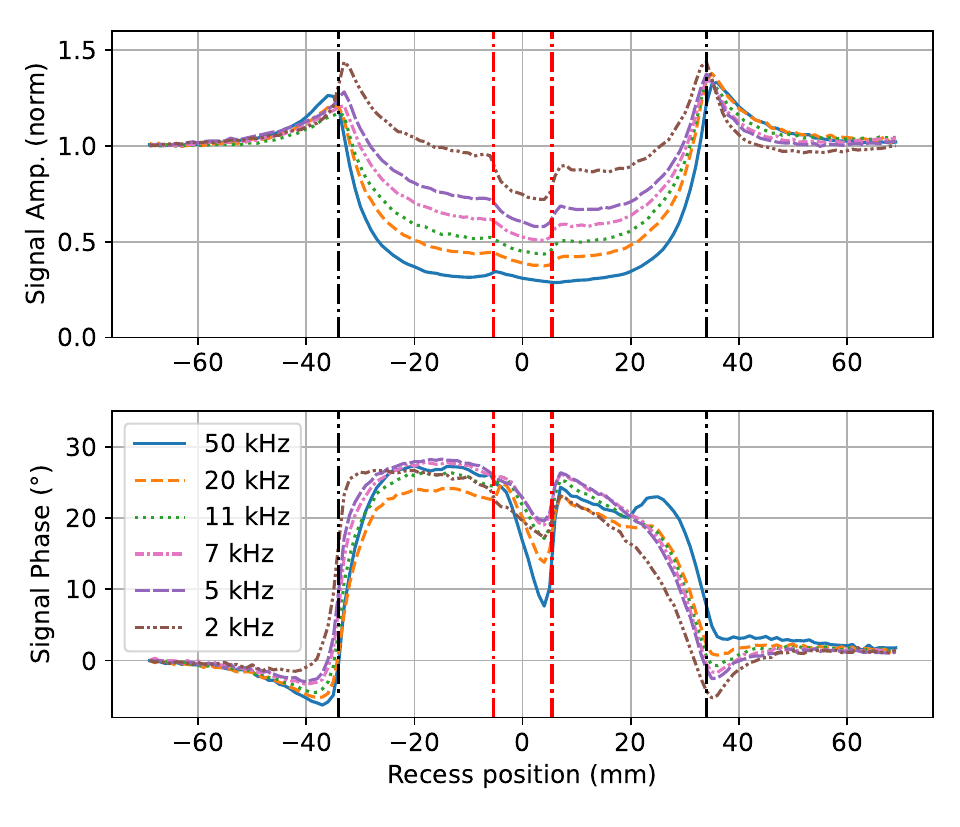}
         \caption{Primary: carbon steel}
         \label{fig:Primary_Steel}
     \end{subfigure}
     \hfill
     \begin{subfigure}[b]{0.49\textwidth}
         \centering
         \includegraphics[width=\textwidth]{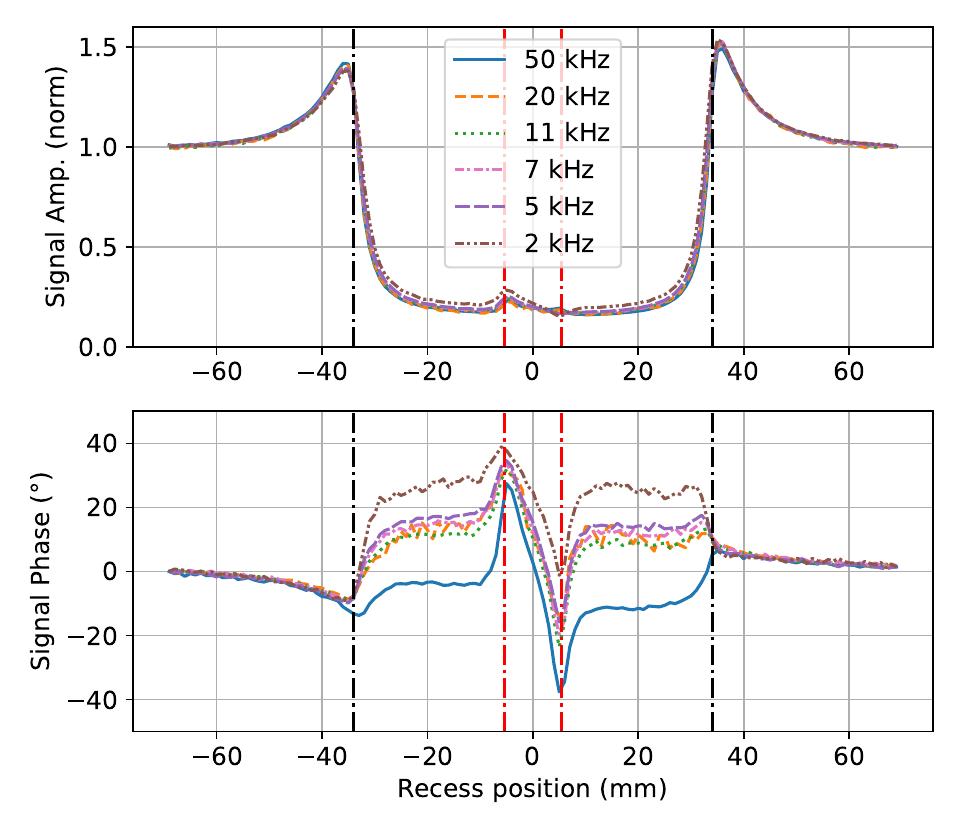}
         \caption{Primary: aluminium}
         \label{fig:Primary_Aluminium}
     \end{subfigure}
    \begin{subfigure}[b]{0.49\textwidth}
         \centering
         \includegraphics[width=\textwidth]{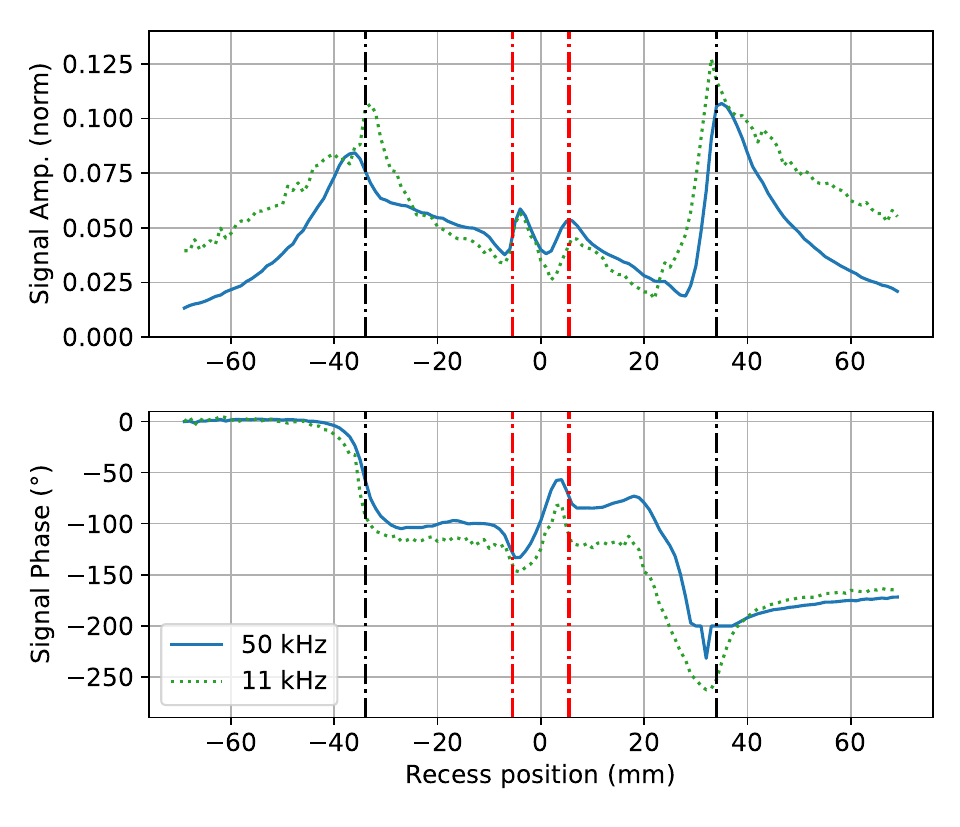}
         \caption{Secondary: carbon steel}
         \label{fig:Secondary_Steel}
     \end{subfigure}
     \hfill
     \begin{subfigure}[b]{0.49\textwidth}
         \centering
         \includegraphics[width=\textwidth]{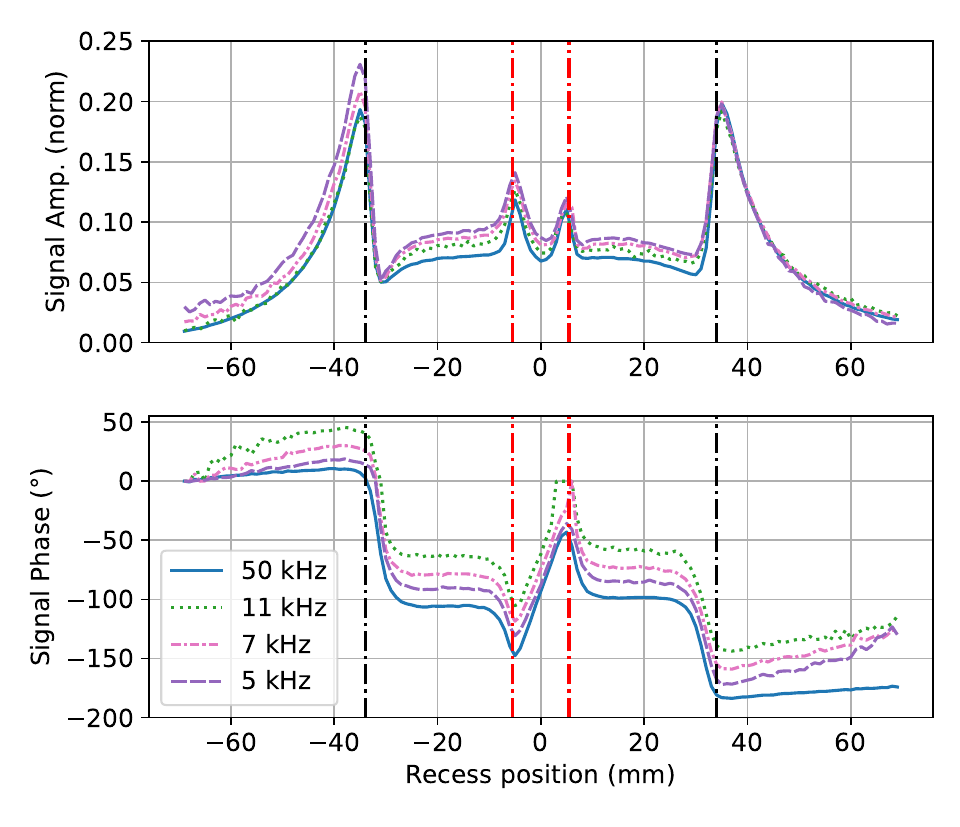}
         \caption{Secondary: aluminium}
         \label{fig:Secondary_Aluminium}
     \end{subfigure}
        \caption{The amplitude and phase (top and bottom row in each figure) of line scans recorded over (a, c) a carbon steel plate and (b, d) an aluminium plate.  These measurements were performed in the (a, b) primary and (c, d) secondary configurations at $\SI{50}{\kilo\hertz}$ (blue, solid), $\SI{20}{\kilo\hertz}$ (orange, dashed), $\SI{11}{\kilo\hertz}$ (green, dotted), $\SI{7}{\kilo\hertz}$ (pink, dash-dot), $\SI{5}{\kilo\hertz}$ (purple, densely dashed) and $\SI{2}{\kilo\hertz}$ (brown, dash-dot-dot). The recorded data in the secondary configurations at $\SI{2}{\kilo\hertz}$ showed the same trends, but were excluded as they were subject to magnetic field drifts. Both plates dimensions $\SI{150}{\milli\meter} \times \SI{150}{\milli\meter} \times \SI{6}{\milli\meter}$ with central $\SI{24}{\milli\meter}$-diameter-$\SI{2.4}{\milli\meter}$-deep-recesses. The plate edges are notated by black dash-dot lines and the recess edges are notated by red dash-dot lines.} 
        \label{fig:PrimarySecondary}
\end{figure*}

Line scans were recorded in Fig.~\ref{fig:PrimarySecondary} for the primary, Figs.~\ref{fig:PrimarySecondary}(a) and (b), and secondary, Figs.~\ref{fig:PrimarySecondary}(c) and (d), configurations for the carbon steel plate, Figs.~\ref{fig:PrimarySecondary}(a) and (c), and the aluminium plate, Figs.~\ref{fig:PrimarySecondary}(b) and (d), in order to study the influence of the secondary field on the measured signal. These measurements were carried out over a range of frequencies to understand the frequency dependence of the signals. As discussed in Sec.~\ref{sec:COMSOLSection}, it can be beneficial to measure different components of the rf field, depending on the object under investigation and the rf frequency. At $\SI{2}{\kilo\hertz}$, i.e., low frequencies, in the steel experimental data, Fig.~\ref{fig:PrimarySecondary}(a), the recess has a notable signature in the amplitude data, mimicking the behaviour of the ferrite and steel COMSOL simulations in Figs.~\ref{fig:COMSOL_Magnetic_Conductor_Mix}(c) and (i), respectively. At higher frequencies, the eddy current driven response begins to dominate over magnetisation effects and the recess signature becomes clearer in the phase response, Fig.~\ref{fig:COMSOL_Magnetic_Conductor_Mix}(f). Since the steel plate has some permanent magnetisation, its movement can cause shifts in the Larmor frequency and the direction of the bias field, even with the external field stabilisation described in Sec.~\ref{sec:ExpSetup}. Additional field correction is achieved by nulling any transverse components of the bias field at each pixel, causing $\textbf{B}_{0}$ to always lie along the $z$- and $y$-axes in the primary and secondary configurations, respectively, and stabilising the Larmor frequency to the desired value, as described in the supplementary material. 
Qualitatively, the amplitude and phase data in Fig.~\ref{fig:COMSOL_Magnetic_Conductor_Mix}(i) is representative of the data recorded in Fig.~\ref{fig:PrimarySecondary}(a). There are more significant differences in the phase data, where experimentally there is a steep phase change at both plate edges, whereas in the simulations there is a smoother change at one edge. It is likely that the differences arise from small misalignments of the primary coil axis, the surface normal of the plate, and the bias field.

Due to the high conductivity of the aluminium plate, $\sim$20~MS/m \cite{Honke2018, elson_meraki_2022}, the signal over the plate drops to roughly the same value for each frequency, as $\omega_{\text{skin}}/(2\pi)\sim350$~Hz for a $\SI{6}{\milli\meter}$ thick object. Measurements and modelling for a lower conductivity sample, Figs.~\ref{fig:COMSOL_Magnetic_Conductor_Mix}(d) and (f), show that reducing operating frequencies does not produce notable features around the defect in the amplitude data, with the greatest signal change visible in the phase data.

\subsection{Secondary configuration}
Figure~\ref{fig:PrimarySecondary} presents line scans over the steel, Fig.~\ref{fig:PrimarySecondary}(c), and aluminium, Fig.~\ref{fig:PrimarySecondary}(d), plates in the secondary (self-compensation) configuration. 
Optimum alignment of the primary rf field with the bias field axis was performed by reducing the amplitude of the magnetometer signal in the absence of the plates. 
The ferromagnetic carbon steel plate produced transverse DC fields that were cancelled at each pixel as described previously.

In contrast to results recorded in the primary configuration at high frequencies, Figs.~~\ref{fig:PrimarySecondary}(a) and (b), the signatures of the recess, with respect to the plate edge, are more visible in the signal amplitudes in the secondary configuration, Figs.~~\ref{fig:PrimarySecondary}(c) and (d). 

Ideally in this measurement geometry, only the $B_{x}$ rf field parallel to the surface of the plate is detectable, making the measurements described in Figs.~\ref{fig:PrimarySecondary}(c) and (d) comparable with those simulated in Figs.~\ref{fig:COMSOL_Magnetic_Conductor_Mix}(h) and (e), respectively. However, this configuration is more susceptible to misalignments between the axes of the bias field, primary field and the object's surface normal. These misalignments may result in a non-zero contribution of the primary field and/or secondary field parallel to the surface normal to the sensor signal. This adds a DC offset (background), affecting contrast/resolution, e.g., due to mixing of signals from different parts of the object.

The data sets for the aluminium plate are relatively frequency-independent in Fig.~\ref{fig:PrimarySecondary}(d). This is consistent with the modelled data for an electrically conductive material, Fig.~\ref{fig:COMSOL_Magnetic_Conductor_Mix}(e). 
Over the homogeneous region of the plate there is a non-zero background amplitude. This comes from the component of the secondary field parallel to the surface normal of the plate.

The permanent magnetism of ferromagnetic objects like steel are challenging to record in the secondary geometry due to the finite stability of the bias magnetic field control. This issue becomes more prominent at smaller bias fields,  when the consequences of geometrical misalignments become more significant. Figure~\ref{fig:PrimarySecondary}(c) shows measurements at 50~kHz and 11~kHz, which generally shows the expected trend of decreasing recess signature amplitudes at low frequencies. A comparison between datasets recorded with steel and aluminium plates at 50~kHz confirms a similar character of the edge and recess signatures. This supports the argument of the steel response being dominated by the electrical conductivity at high frequencies.

It should be noted that the edge signal outside of the plate at 11~kHz [green dotted line in Fig.~\ref{fig:PrimarySecondary}(c)] is elevated with respect to 50~kHz. Inaccuracies in nulling the transverse DC magnetic fields produced by the steel plate result in a non-zero signal recorded outside the plate at 11 kHz. One method to reduce the influence of changing transverse DC magnetic fields is to drive an rf two-photon transition with one high- and one low-frequency rf fields, whose sum equals the Larmor frequency \cite{Geng2021,Maddox2023}. This enables operation at a high bias magnetic field, while maintaining a low rf frequency for the MIT measurement. Initial tests in this configuration show that the amplitude response for steel is similar to what is expected from the simulation in Fig.~\ref{fig:COMSOL_Magnetic_Conductor_Mix}(h). Explorations of the two-photon signal response are beyond the scope of this work, but is currently being studied and will be reported on in future publications.

Visual analysis of Fig.~\ref{fig:PrimarySecondary} shows that it is favourable to measure ferromagnetic magnetically permeable objects in the primary geometry, where the greatest change in the recess signature is seen at low frequencies and is supported by the simulations shown in Fig.~\ref{fig:COMSOL_Magnetic_Conductor_Mix}, whereas purely conductive objects are better suited to the secondary configuration. The different configurations for different object compositions are required because of the intrinsic response signal generation, as well as the difficulties in stabilising the bias field.

\section{Conclusions}
Valid interpretations of the measurement results require an understanding of the sensor's properties. We explored the rf atomic magnetometer's sensitivity to the polarisation of oscillating magnetic fields. This issue becomes relevant when the measured field is produced by sources with different characteristics. Such a scenario is realized in MIT measurements over, for example, stainless steel, where the secondary fields are produced by eddy current and magnetisation contributions, whose phases are different with respect to the driving field. This results in the total field changing its polarisation across the studied object. The change of the field polarisation can affect the signal magnitude $R$. In tomographic measurements the amplitude of the defect signature is a measure of the defect depth. Consequently, rf field polarisation variations could be misinterpreted as a change in a defect's depth. This could be avoided by measurements performed with opposite directions of the bias field, which makes the sensor sensitive to orthogonal circular polarisations.

By modelling responses for objects with different compositions and simultaneously monitoring individual field components, we have shown the ability to optimise the defect/object detection. We have demonstrated that their signatures could be visible in different field components depending on the object's electrically conductive and magnetically permeable properties. Additionally, visibility of the defect signatures could be improved by tuning the operation frequency. Analysis of the frequency dependence also provides an indicator of the object's dominant property, i.e., its composition. 

In general, experimental inductive measurements were compared and validated with modelling results from COMSOL. This creates the opportunity for relatively quick explorations of limits of the discussed technique in various application scenarios, e.g., detecting defects in objects at close range (NDT) or object detection at long range (for example in underground radar and metal detector systems), with a consequent optimisation of the sensing scheme. 

\section{Acknowledgements}
We acknowledge the support of the UK government Department for Science, Innovation and Technology through the UK national quantum technologies program. We would like to thank R. Hendricks for critically reading this manuscript.

\section{Data Availability Statement}
The data that support the findings of this study are available from the corresponding author upon reasonable request.


\end{document}


\setcounter{equation}{0}
\renewcommand\theequation{A.\arabic{equation}}
\renewcommand\thefigure{A.\arabic{figure}}
\maketitle

\section{Solving the Bloch equations in a 2D rf field}

The output of the atomic magnetometer can be understood mathematically by solving the Bloch equations for the ensemble of caesium spins with collective atomic spin $\textbf{J}$ \cite{rushton_2022, jensen_2019}. Identifying $\textbf{B}=\textbf{B}_{0}+\textbf{B}_{\text{rf}}(t)$, where $\textbf{B}_{\text{rf}}(t) = B_{x}(t)\hat{\textbf{x}} + B_{y}(t)\hat{\textbf{y}}$, we write the Bloch equation in the laboratory frame as
\begin{equation}
\frac{d\textbf{J}}{dt} = \textbf{J} \times \gamma\textbf{B} + \textbf{J}_{\text{max}} - (R_{p} + \Gamma)\textbf{J},
\label{eq:dJdt_SpinsInStaticandOscillatingMagneticFields}
\end{equation}
where $\gamma=3.5$Hz/nT denotes caesium's gyromagnetic ratio \cite{steck_cs}, $\textbf{J}_{\text{max}} = R_{p} J_{\text{max}}\hat{\textbf{z}}$ is the net momentum of a fully-polarised caesium ensemble (subject to pumping rate $R_p$ along the bias field), and $\Gamma$ describes relaxation mechanisms. For ease of computation, we rewrite our rf fields in the form $B_{x}(t) = B_{\text{x,c}} \cos (\omega_{\text{rf}} t) + B_{\text{x,s}} \sin (\omega_{\text{rf}} t)$ with $B_{\text{x,c}}=|B_{x}|\cos(\phi_{x})$ and $B_{\text{x,s}}=|B_{x}|\sin(\phi_{x})$ [and equivalent for $B_{y}(t)$]. According to vector transformation rules \cite{ThorntonBook}, we may rewrite the Bloch equation in a new frame, denoted by primes, which rotates about $\hat{\textbf{z}}$ with frequency and direction $\boldsymbol{\omega}_{\text{rf}} = -\omega_{\text{rf}}\ \hat{\textbf{z}}$

\begin{equation}
\frac{d\textbf{J}'}{dt} = \textbf{J}' \times \gamma\left(\textbf{B}' + \frac{\boldsymbol{\omega}_{\text{rf}}}{\gamma}\right)+ \textbf{J}_{\text{max}} - (R_{p} + \Gamma)\textbf{J}'.
\label{eq:dJdt_SpinsInRotFrame}
\end{equation}
We solve this set of three equations according to the steady state $d\textbf{J}'/dt = 0$, writing each explicitly as

\begin{equation}
    \begin{aligned}
    \frac{dJ'_x}{dt} &= -\delta\omega J'_{x} -\Delta_{\text{rf}}J'_{y} -\gamma J'_z B'_y =0,\\
    \frac{dJ'_y}{dt} &= \Delta_{\text{rf}}J'_x -\delta\omega J'_{y} +\gamma J'_z B'_x =0,\\
    \frac{dJ'_z}{dt} &=  \gamma J'_x B_y' - \gamma J_y' B_x' -\delta\omega J_z'+ R_p J_{\text{max}}=0,\\
    \end{aligned}
    \label{linearODESystem}
\end{equation}
where $\delta\omega = R_{p} + \Gamma$ and $\Delta_{\text{rf}} = \omega_\text{rf}-\gamma B_0$. Consider $B_x$ and $B_y$ in the rotating frame as $B'_x$ and $B'_y$, making the change of basis
 \begin{equation}
 \begin{aligned}
     \hat{{\textbf{x}}} &= \cos{\omega_{\text{rf}} t} \ \hat{\textbf{x}}' + \sin{\omega_{\text{rf}} t} \ \hat{\textbf{y}}',\\  
     \hat{\textbf{y}} &= -\sin{\omega_{\text{rf}} t}  \ \hat{\textbf{x}}' + \cos{\omega_{\text{rf}} t} \ \hat{\textbf{y}}'.\\
 \end{aligned}
 \end{equation}
The fields in this frame will have time-dependent terms proportional to $\cos^2{\omega_{\text{rf}} t}$, $\sin^2{\omega_{\text{rf}} t}$ and $\sin{\omega_{\text{rf}} t}\cos{\omega_{\text{rf}} t}$. 
We average over each of these in the limit $T \gg 2\pi/\omega_{\text{rf}}$ to eliminate the time-dependence. Consider the following quantities
\begin{equation}
    \begin{aligned}
        \langle \sin^2(\omega_{\text{rf}} t) \rangle &= \frac{1}{T}\int_0^T \frac{1-\cos(2\omega_{\text{rf}} t)}{2} dt \sim \frac{1}{2},\\
        \langle \cos^2(\omega_{\text{rf}} t) \rangle &= \frac{1}{T}\int_0^T \frac{1+\cos(2\omega_{\text{rf}} t)}{2} dt \sim \frac{1}{2},\\
        \langle \sin(\omega_{\text{rf}} t)\cos(\omega_{\text{rf}} t) \rangle &= \frac{1}{T}\int_0^T \frac{\sin (2\omega_{\text{rf}}t)}{2} dt \sim 0. \\
    \end{aligned}
\end{equation}
In particular, where each rf field can be thought of as the sum of co- and counter-rotating fields, in the rotating frame the co-rotating field appears stationary, while the counter-rotating field rapidly averages to zero \cite{Ledbetter2007}. This manifests the rotating wave approximation to give

\begin{equation}
    \begin{aligned}
     \langle B_x' \rangle &=  \frac{1}{2}\left(B_{\text{x,c}} - B_{\text{y,s}}\right), \\
     \langle B_y' \rangle &=  \frac{1}{2}\left(B_{\text{y,c}} + B_{\text{x,s}}\right). \\
    \end{aligned}
\end{equation}
With hindsight, we also write

\begin{equation}
    \begin{aligned}
     \langle B_x' \rangle^2 &= \frac{1}{4}\left(B^2_{\text{x,c}} + B^2_{\text{y,s}}-2B_{\text{x,c}}B_{\text{y,s}}\right), \\
     \langle B_y' \rangle^2 &=  \frac{1}{4}\left(B^2_{\text{x,s}} + B^2_{\text{y,c}}+2B_{\text{x,s}}B_{\text{y,c}}\right). \\
    \end{aligned}
\end{equation}
To solve Eq. \ref{linearODESystem}, we can write the system of equations as the matrix $M$

\begin{equation}
\begin{bmatrix}
-\delta\omega  & -\Delta_{\text{rf}} & -\gamma \langle B_y' \rangle\\ 
\Delta_{\text{rf}} & -\delta\omega &  \gamma \langle B_x' \rangle\\
\gamma \langle B_y' \rangle & -\gamma \langle B_x' \rangle & -\delta\omega \\ 
\end{bmatrix}
 \begin{bmatrix} J'_{x} \\ J'_{y}\\J'_{z} \end{bmatrix} =  \begin{bmatrix}  0 \\ 0 \\ -R_p J_{\text{max}} 
 \end{bmatrix}.
 \label{matsystem}
\end{equation}
The inverse of this matrix follows 
\begin{equation}M^{-1} = 
\frac{1}{det(M)}
\begin{bmatrix}
\delta\omega^2 +\gamma^2 \langle B_x' \rangle^2 & -\Delta_{\text{rf}}\delta\omega + \gamma^2 \langle B_y' \rangle \langle B_x' \rangle& -\gamma\Delta_{\text{rf}}\langle B_x' \rangle-\gamma\delta\omega \langle B_y' \rangle\\ 
 -\Delta_{\text{rf}}\delta\omega -\gamma^2 \langle B_x' \rangle \langle B_y' \rangle& \delta\omega^2 +\gamma^2 \langle B_y' \rangle^2 & \gamma\delta\omega \langle B_x' \rangle - \gamma \Delta_{\text{rf}} \langle B_y' \rangle \\
\gamma\Delta_{\text{rf}}\langle B_x' \rangle-\gamma\delta\omega \langle B_y' \rangle & -\gamma \delta\omega \langle B_x' \rangle - \gamma \Delta_{\text{rf}} \langle B_y' \rangle &\delta\omega^2 + \Delta_{\text{rf}}^2   \\ 
\end{bmatrix},
\end{equation}
\begin{equation}
    det(M) = -\delta\omega^3 -\delta\omega\Delta_{\text{rf}}^2 - \delta\omega\gamma^2(\langle B_x' \rangle^2 + \langle B_y' \rangle^2).
\end{equation}
Applying the inverse to Eq. \ref{matsystem} and substituting in the averaged magnetic fields gives the following solutions
\begin{align}
    J'_{x} &= -J_{\text{ss}}\frac{\gamma(\Delta_{\text{rf}}[B_{\text{\text{x,c}}}-B_{\text{y,s}}]  + \delta \omega[B_{\text{x,s}}+B_{\text{y,c}}] )/2}{\delta \omega^{2} + \gamma^{2}  (B_{\text{\text{x,c}}}^{2} + B_{\text{x,s}}^{2} + B_{\text{y,c}}^{2} + B_{\text{y,s}}^{2}+2B_{\text{x,s}}B_{\text{y,c}}-2B_{\text{x,c}}B_{\text{y,s}})/4 + \Delta_{\text{rf}}^{2}},
\label{eq:Jx'}\\
J'_{y} &= J_{\text{ss}}\frac{\gamma (\delta \omega [B_{\text{\text{x,c}}}-B_{\text{y,s}}]- \Delta_{\text{rf}}[B_{\text{x,s}}+B_{\text{y,c}}] ) /2}{\delta \omega^{2} + \gamma^{2} (B_{\text{\text{x,c}}}^{2} + B_{\text{x,s}}^{2} + B_{\text{y,c}}^{2} + B_{\text{y,s}}^{2} + 2B_{\text{x,s}}B_{\text{y,c}}-2B_{\text{x,c}}B_{\text{y,s}})/4 + \Delta_{\text{rf}}^{2}},
\label{eq:Jy'}\\
J'_{z} &= J_{\text{ss}} \frac{\delta \omega^{2} + \Delta_{\text{rf}}^{2}}{\delta \omega^{2} + \gamma^{2} (B_{\text{\text{x,c}}}^{2} + B_{\text{x,s}}^{2} + B_{\text{y,c}}^{2} + B_{\text{y,s}}^{2}+2B_{\text{x,s}}B_{\text{y,c}}-2B_{\text{x,c}}B_{\text{y,s}})/4 + \Delta_{\text{rf}}^{2}},
\label{eq:Jz'}
\end{align}
with $J_{\text{ss}} = R_p J_{\text{max}}/\delta\omega$. Demodulating this signal, we identify the in- and out-of-phase components as $X\propto J_{y}'$ and $Y\propto J_{x}'$. Moreover, for an on-resonance rf field ($\Delta_{\text{rf}}=0$) in the limit $\delta\omega^2 \ll \gamma^2(\langle B_x' \rangle^2 + \langle B_y' \rangle^2)$, we find
\begin{equation}
    X\propto B_{\text{x,c}}-B_{\text{y,s}},
    \label{eq:LockinX}
\end{equation}
\begin{equation}
   Y\propto -(B_{\text{x,s}}+B_{\text{y,c}}). 
   \label{eq:LockinY}
\end{equation}
From $X$ and $Y$, the signal amplitude and signal phase of the rf atomic magnetometer can be calculated as
\begin{equation}
    \text{Signal Amp.} = \sqrt{X^{2}+Y^{2}}
    \label{eq:SignalAmp}
\end{equation}
and
\begin{equation}
    \text{Signal Phase} = \arctan(\frac{Y}{X}).
\label{eq:SignalPhase}
\end{equation}

\section{Describing polarisation with Stokes parameters}

Figure 3(c) in the main text describes the polarisation of two orthogonal rf fields according to Stokes parameters $S_{0,1,2,3}$, normalised such that $s_1^2 + s_2^2 + s_3^2 = 1$, where $s_{1,2,3}=S_{1,2,3}/S_0$. Respectively, these define the total field's energy and projections onto vertical/horizontal linear, $\pm45\degree$ linear and left/right-handed circular polarisation axes. These definitions (in terms of initial rf fields) and their relation to the Poincar\'e sphere are made explicit in \cite{Walkenhorst2020}, as is a comment on the difficulty of representing polarisation evolution on a sphere in publication or display. As the coils sweeps through a phase difference of $360\degree$, we see the total field cycle between perfect $\pm45\degree$ linear polarisations (at $\Delta\Theta = 0\degree, 180\degree$ and $360\degree$) and perfect left and right-handed circular polarisations (at $\Delta\Theta = 90\degree$ and $270\degree$, respectively). 


We can determine how our measured signal amplitude, denoted here as $R$, depends on the Stokes parameters by redefining the latter with respect to our orthogonal rf fields \cite{Walkenhorst2020} 
\begin{align}
    S_{0}&~=B_{x}^{2}+B_{y}^{2},\\
    S_{1}&~=B_{x}^{2}-B_{y}^{2},\\
    S_{2}&~=2a_{1}a_{2}\cos(\phi_{y}-\phi_{x}),\\
    S_{3}&~=2a_{1}a_{2}\sin(\phi_{y}-\phi_{x}),
\end{align}
where $a_{1}=\text{abs}[B_{x}]$ and $a_{2}=\text{abs}[B_{y}]$.
Using Eqs.~\ref{eq:LockinX} and \ref{eq:LockinY},
\begin{align}
    R^{2}&~=X^{2}+Y^{2}\\&~=B_{x}^{2}+B_{y}^{2}+2B_{x}B_{y}(\sin\phi_{x}\cos\phi_{y}-\cos\phi_{x}\sin\phi_{y}).
\end{align}
Using the trigonometric identity $-\sin(\phi_{y}-\phi_{x})=\sin\phi_{x}\cos\phi_{y}-\cos\phi_{x}\sin\phi_{y}$, we obtain

\begin{equation}
    R^{2}=S_{0}-2B_{x}B_{y}\sin(\phi_{y}-\phi_{x}),
\end{equation}
which simplifies to
\begin{equation}
    R=\sqrt{S_{0}-S_{3}}.
\end{equation}
Figure 6 in the main text visualises this relationship, plotting $R$, $S_0$ and $S_3$ for COMSOL data in the primary geometry with $\mu_{r}=80$, $\sigma=3$~MS/m.

\begin{figure}[b!]
    \centering
     \includegraphics[width=0.8\textwidth]{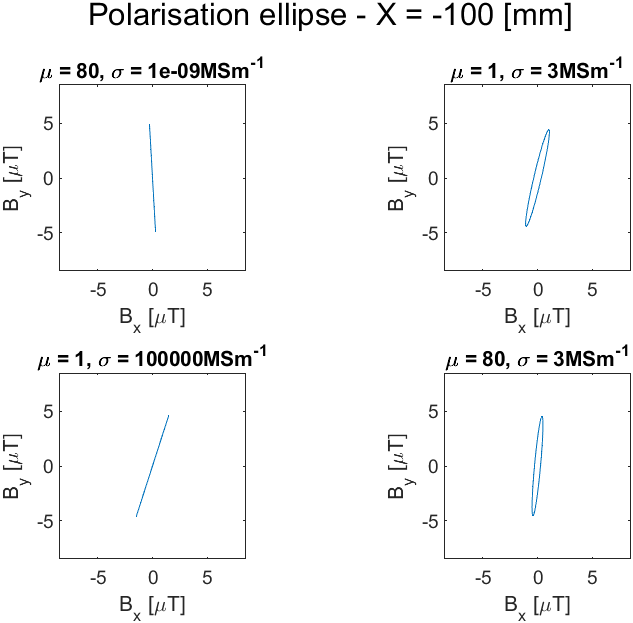}
    \caption{A snapshot at $x=-100$~mm of an animation demonstrating the change in polarisation for various samples for COMSOL simulations in the primary geometry. This GIF can be viewed in the attached files.}
    \label{fig:gif_insert}
\end{figure}

Figure~\ref{fig:gif_insert} shows polarisation ellipses plotted for four different permeability-conductivity combinations in the primary configuration, including (a) $\mu_{r}=$~80, $\sigma\sim0$, (b) $\mu_{r}=$~1, $\sigma=$~3~MS/m, (c) $\mu_{r}=$~1, $\sigma=$10$^{5}$~MS/m, (d) $\mu_{r}=80$, $\sigma=3$~MS/m. For the samples which are only (a) magnetically permeable and (c) highly electrically conductive, it can be seen that the rf fields remain only linear when the plate is scanned under the rf coil. However, the rf field changes polarisation across the other two plates.

\section{Frequency dependence of inductive measurements for electrically conductive and magnetically permeable samples}
\label{sec:SupplementaryFrequencyDependence}
\begin{figure}[h]
    \centering
    \includegraphics[width=0.6\linewidth]{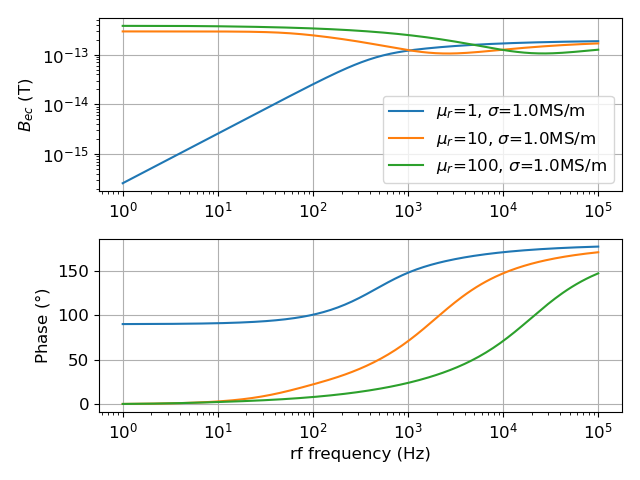}
    \caption{Amplitude and phase of the on-axis secondary magnetic field induced in spheres with a conductivity $\sigma=1$~MS/m and magnetic permeabilities $\mu_{r}=1,10,100$ as a function of rf frequency using Eq.~\ref{eq:bidinosti}.}\label{fig:COMSOLvsTHEORYPermeability}
\end{figure}

The frequency dependence of inductive measurements when an electrically conductive and/or magnetically permeable sample is placed in an rf field can be understood using the theory derived in \cite{Bidinosti2007, Honke2018}, where the on-axis secondary magnetic field $B_{\text{ec}}$ produced by a sphere with a radius $a$, electrical conductivity $\sigma$ and magnetic permeability $\mu=\mu_{r}\mu_{0}$ placed in a uniform rf field is calculated to be
\begin{equation}
\label{eq:bidinosti}
    B_{\text{ec}} = \frac{2\mu_{0}m}{4\pi r^{3}},
\end{equation}
where $r$ is the distance from the excitation coil to the sphere and the distance from the sphere to the sensing position and $m$ is the induced magnetic moment in the sphere
\begin{equation}
    m = \frac{2\pi a^{3}B_{1}}{\mu_{0}}\frac{[2(\mu-\mu_{0})j_{0}(ka)+(2\mu+\mu_{0})j_{2}(ka)]}{[(\mu+2\mu_{0})j_{0}(ka)+(\mu-\mu_{0})j_{2}(ka)]},
\end{equation}
where $B_{1}$ is the primary field at the position of the sphere, $k=\sqrt{\mu\epsilon\omega^{2}+i\mu\sigma\omega}$ is the propagation constant and
\begin{equation}
    j_{0}(ka)=\frac{\sin(ka)}{ka},
\end{equation}
\begin{equation}
    j_{2}(ka)=(\frac{3}{(ka)^{3}}-\frac{1}{ka})\sin(ka)-\frac{3}{(ka)^{2}}\cos(ka)
\end{equation}
are the $j_{n}$ spherical Bessel functions. Using some arbitrary parameters of $B_{1}=1.5$~nT, $a=5$~cm, $r=1$~m, $\mu_{r}=(1, 10, 100)$, $\sigma=1$~MS/m and varying the rf frequency from 1~Hz to 100~kHz, the amplitude and phase of the secondary magnetic field $B_{\text{ec}}$ for each sphere is plotted in Fig.~\ref{fig:COMSOLvsTHEORYPermeability}. At high frequencies, all spheres tend to the same secondary magnetic field limit, which is governed by the radius $a$ of the sphere and the distance $r$ of the sphere from the excitation coil. The spheres with $\mu_{r}=10,100$ tend to the values of the sphere with $\mu_{r}=1$, as at high frequencies eddy currents dominate over magnetisation. At low frequencies, the spheres with $\mu_{r}>1$ produce large secondary magnetic fields, due to magnetisation dominating over eddy current effects at low frequencies.

\section{Controlling the transverse and bias magnetic fields}
\begin{figure*}
     \centering
     \begin{subfigure}[b]{0.329\textwidth}
         \centering
         \includegraphics[width=\textwidth]{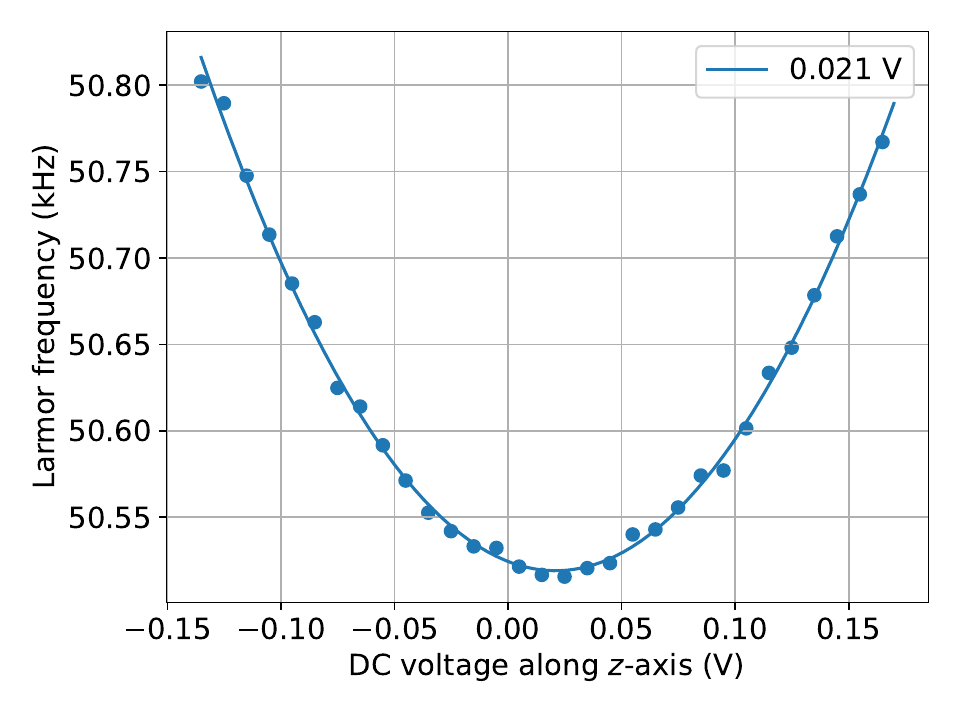}
         \caption{}
         \label{fig:}
     \end{subfigure}
     \hfill
     \begin{subfigure}[b]{0.329\textwidth}
         \centering
         \includegraphics[width=\textwidth]{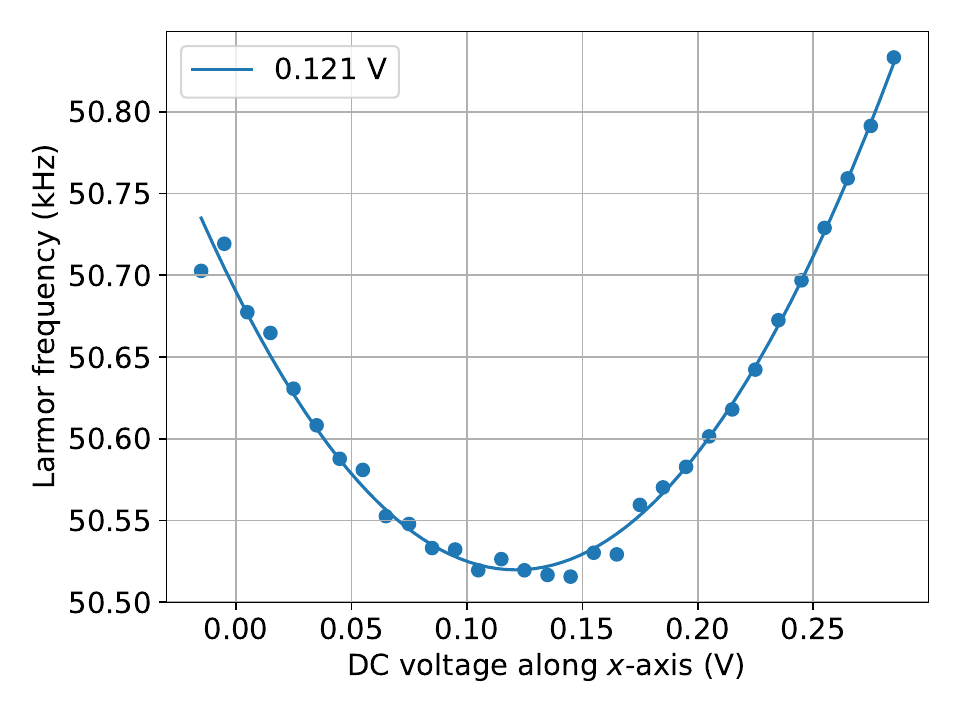}
         \caption{}
         \label{fig:}
     \end{subfigure}
     \hfill
     \begin{subfigure}[b]{0.329\textwidth}
         \centering
         \includegraphics[width=\textwidth]{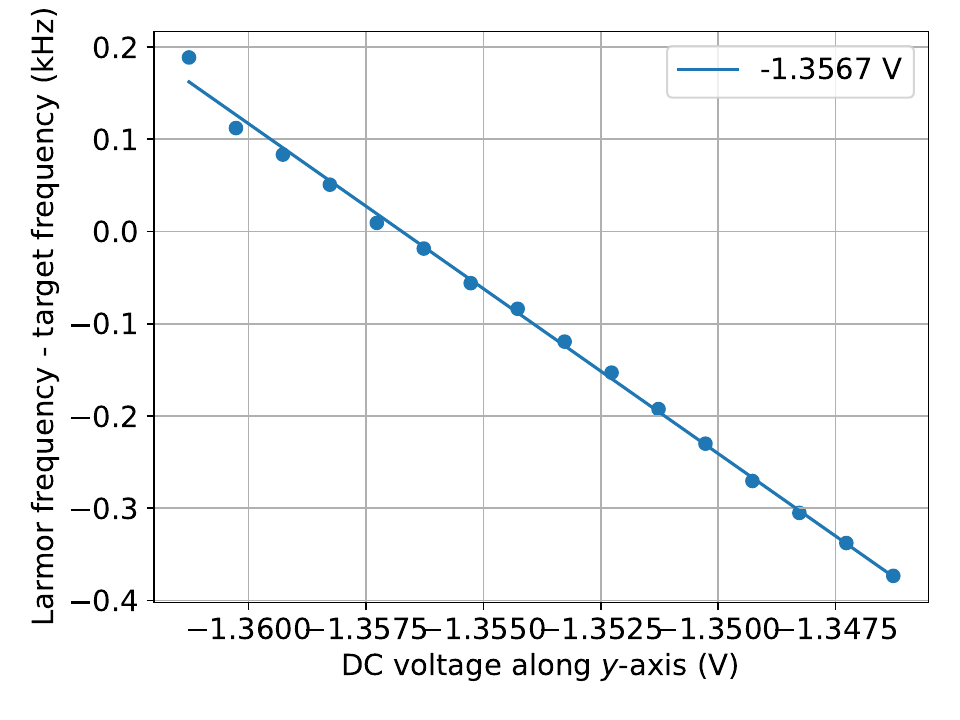}
         \caption{}
         \label{fig:}
     \end{subfigure}
        \caption{Magnetic field stabilisation technique. At each pixel, the DC field was varied first along the two transverse fields, (a) and (b), then along the bias field, (c). This allowed for the DC fields at each pixel to be controlled. The voltage in each legend is the voltage applied to each coil.}
        \label{fig:MagneticFieldStabilisation}
\end{figure*}

Due to the ferromagnetism of the steel plate, field control was performed at each pixel. To obtain a bias field of $\omega_{L}/(2\pi)=50$~kHz along the $y$-axis, the first transverse field along the $z$-axis was varied, leading to the example dataset in Fig.~\ref{fig:MagneticFieldStabilisation}(a). The Larmor frequency of the resonance signal was recorded for each applied DC voltage along the $z$-axis. The minimum of the quadratic fit, equal to 0.021~V, was used as the DC value to null the static field along the $z$-axis. A similar procedure was used along the other transverse axis, where a voltage of 0.121~V was required to null the fields along this axis. Finally, the bias magnetic field was varied and the zero-crossing (Larmor frequency - target frequency) was used to set the Larmor frequency (equal to -1.3567~V).

